\documentclass[11pt]{article}

\usepackage{pstricks}
\usepackage{pst-plot}
\usepackage{pst-node}
\usepackage{amsmath}
\usepackage{pst-grad}
\usepackage{amssymb}
\usepackage{xspace}
\usepackage{pst-math}
\usepackage{pst-xkey}
\usepackage{pst-coil}
\usepackage{pst-3dplot}
\usepackage[T1]{fontenc}
\usepackage{ae}
\usepackage[ansinew]{inputenc}
\usepackage{graphics}
\usepackage{color}
\usepackage{float}
\usepackage{graphicx}

\newrgbcolor{darkred}{0.8 0 0}
\newrgbcolor{darkerred}{0.7 0 0}
\newrgbcolor{darkestred}{0.5 0 0}
\newrgbcolor{darkgreen}{0 0.5 0.5}
\newrgbcolor{darkergreen}{0 0.6 0}
\newrgbcolor{darkestgreen}{0 0.5 0}
\newrgbcolor{darkblue}{0 0 0.6}
\newrgbcolor{darkestblue}{0 0 0.5}
\newrgbcolor{redgreen}{0.8 0.8 0}
\newrgbcolor{bluegreen}{0 0.4 0.7}
\newrgbcolor{bluewhite}{0 0.9 1}
\newrgbcolor{lightgreen}{0.6 1 0.6}
\newrgbcolor{lightred}{1 0.8 0.8}
\newrgbcolor{lightyred}{1 0.4 0.4}
\newrgbcolor{lightyblue}{0.5 0.5 1}
\newrgbcolor{lightblue}{0.8 1 1}
\newrgbcolor{lilas}{0.5 0.3 0.7}
\newrgbcolor{lilaswhite}{1 0.8 1}
\newrgbcolor{redy}{1 0.6 0.6}
\newrgbcolor{bluish}{0.6 0.6 1}
\newrgbcolor{sand}{0.9 0.9 0.7}
\newrgbcolor{brown}{0.4 0.4 0}
\newrgbcolor{blue1}{0 0.35 0.65}
\newrgbcolor{blue2}{0 0.4 0.6}
\newrgbcolor{blue3}{0 0.45 0.55}
\newrgbcolor{blue4}{0 0.5 0.5}
\newrgbcolor{blue5}{0 0.55 0.45}
\newrgbcolor{blue6}{0 0.6 0.4}
\newrgbcolor{blue7}{0 0.65 0.35}

\begin{document}

\title{Structure and causality relations in a global network of financial companies.}

\author{Leonidas Sandoval Junior \\ \\ Insper, Instituto de Ensino e Pesquisa}

\maketitle

\begin{abstract}
This work uses the stocks of the 197 largest companies in the world, in terms of market capitalization,  in the financial area in the study of causal relationships between them using Transfer Entropy, which is calculated using the stocks of those companies and their counterparts lagged by one day. With this, we can assess which companies influence others according to sub-areas of the financial sector, which are banks, diversified financial services, savings and loans, insurance, private equity funds, real estate investment companies, and real estate trust funds. We also analyzed the causality relations between those stocks and the network formed by them based on this measure, verifying that they cluster mainly according to countries of origin, and then by industry and sub-industry. Then we collected data on the stocks of companies in the financial sector of some countries that are suffering the most with the current credit crisis: Greece, Cyprus, Ireland, Spain, Portugal, and Italy, and assess, also using transfer entropy, which companies from the largest 197 are most affected by the stocks of these countries in crisis. The intention is to map a network of influences that may be used in the study of possible contagions originating in those countries in financial crisis.
\end{abstract}

\section{Introduction}

In his speech delivered at the Financial Student Association in Amsterdam, in 2009, Andrew G. Haldane (2009), Executive Director of Financial Stability of the Bank of England, called for a rethinking of the financial network, that is the net formed by the connections between banks and other financial institutions. He warned that, in the last decades, this network had become more complex and less diverse, and that these facts may have led to the crisis of 2008.

According to him, it was the belief of theoreticians and practitioners of the financial market that connectivity between financial companies meant risk diversification and dispersion, but further studies showed that networks of certain complexity exhibit a robust but fragile structure, where crises may be dampened by sharing a shock among many institutions, but where they may also spread faster and further due to the connections between companies. Other issue to be considered was the fact that some nodes in the financial network were very connected to others, while some were less connected. The failure of a highly connected node could, thus, spread a small crisis to many other nodes in the network. Another factor was the small-world property of the financial network, where one company was not very far removed from another, through relations between common partners, or common partners of partners.

Such connected network was also more prone to panic, tightening of credit lines, and distress sales of assets, some of them caused by uncertainties about who was a counterpart to failing companies. Due to some financial innovations, risk was now shared among many parties, some of them not totally aware of all the details of a debt that was sectorized, with risk being decomposed and then reconstituted in packages that were then resold to other parties. This made it difficult to analyze the risk of individual institutions, whose liabilities were not completely known even to themselves, since they involved the risks of an increasingly large number of partners.

The other important aspect, the loss of diversity, increased when a large number of institutions adopted the same strategies in the pursuit of return and in the management of risk. Financial companies were using the same models and using the same financial instruments, with the same aims.

In the same speech, Haldane pointed at some directions that could improve the stability of the financial network. The first one was to map the network, what implied the collection, sharing and analysis of data. This analysis needed to include techniques that didn't focus only on the individual firms, like most econometric techniques do, but also on the network itself, using network techniques developed for other fields, like ecology or epidemiology. The second was to use this knowledge to properly regulate this network. The third was to restructure the financial network, eliminating or reinforcing weak points. All these need a better understanding of the connections between financial institutions and how these connections influence the very topology of the financial network.

This article contributes to the first direction pointed by Haldane, that of understanding the international financial network. We do it by calculating two types of networks based on the daily returns of the stocks of the 197 largest financial companies across the world in terms of market capitalization that survive a liquidity filter. These include not just banks, but also diversified financial services, insurance companies, one investment company, a private equity, real estate companies, REITS (Real Estate Investment Trusts), and savings \& loans institutions. We use the daily returns in order to build the networks because we believe that the price of a stock encodes a large amount of information about the company to which it is associated that goes beyond the information about the assets and liabilities of the company. Also, we believe that it is more interesting to study the effects of stock prices on other stock prices, as in the propagation of a financial crisis, rather than the spreading of defaults, since defaults are are events that are usually avoided by injecting external capital into banks.

The first network is built on the correlations between the log-returns of those equities, and the second one is built using Transfer Entropy, a measure first developed in information science. The first network is an undirected one, which expresses the common movements of stocks, and the second one is a directed one, which reveals causality relations between equities. Both networks are used in order to determine which are the most central nodes, according to diverse centrality criteria. We also enlarge the original network obtained by Transfer Entropy to include the most liquid stocks belonging to financial companies in some European countries that have been receiving much attention recently due to the fact that they are facing different degrees of economic crises, and determine who are the major financial companies in the world that are most affected by price movements of those stocks, and which of those stocks belonging to countries in crisis are the most influent ones.

There is an extensive literature on the propagation of shocks in networks of financial institutions, and describing all the published works in this subject is beyond the scope of this article. So, we shall here only cite the article that is considered the seminal work in networks of financial institutions and some review articles in the field. Most of the works in this field can be divided into theoretical and empirical ones, most of them considering networks of banks where the connections are built on the borrowing and lending between them. In most theoretical works, networks are built according to different topologies (random, small world, or scale-free), and the propagation of defaults is studied on them. The conclusions are that small world or scale-free networks are, in general, more robust to cascades (the propagation of shocks) than random networks, but they are also more prone to propagations of crises if the most central nodes (usually, the ones with more connections) are not themselves backed by sufficient funds. Most empirical works are also based on the structure derived from the borrowing and lending between banks, and they show that those networks exhibit a core-periphery structure, with few banks occupying central, more connected positions, and others populating a less connected neighbourhood. Those articles showed that this structure may also lead to cascades if the core banks are not sufficiently resistant, and that the network structures changed considerably after the crisis of 2008, with a reduction on the number of connected banks and a more robust topology against the propagation of shocks.

The work that is considered the first that deals with the subject is the one of Allen and Gale (2000), where the authors modeled financial contagion as an equilibrium phenomenon, and concluded that equilibrium is fragile, that liquidity shocks may spread through the network, and that cascade events depend on the completeness of the structure of interregional claims between banks. In their model, they used four different regions, which may be seen as groups of banks with some particular specializations. They focused in one channel of contagion, which is the overlapping claims that different regions or sectors of the banking system have on one another. According to them, another possible channel of contagion that is not being considered is incomplete information among agents. As an example, the information of a shock in one region may create a self-fulfilling shock in another region if that information is used as a prediction of shocks in other regions. Another possible channel of contagion is the effect of currency markets in the propagation of shocks from one country to another. In their results, the spreading of a financial crisis depends crucially on the topology of the network. A completely connected network is able to absorb shocks more efficiently, and a network with strong connections limited to particular regions which are not themselves well connected is more prone to the dissemination of shocks.

Later, Allen and Babus (2009)  made a review of the progress of the network approach to the propagation of crises in the financial market. They concluded that there is an urgent need for empirical work that maps the financial network, so that the modern financial systems may be better understood, and that a network perspective would not only account for the various connections within the financial sector or between the financial sector and other sectors, but also would consider the quality of such links. Upper (2011) made a survey of a diversity of simulation methods that have been used with a variety of financial data in order to study contagion in financial networks, and made a comparison between the various methods used.

This article is organized as follows. Section 2 explains the data used in the article and some of the methodology. Section 3 uses the correlations between stocks in order to exemplify some of the techniques to be used for Transfer Entropy, but yet in a more familiar background. Section 4 explains Transfer Entropy and uses it in order to study the causality relations between the stocks of financial institutions. That section also highlights which are the most central stocks according to different centralities criteria. Section 5 studies the relationships between countries in crisis in Europe with the largest financial institutions, analyzing which stocks are more affected by movements in the stocks belonging to those countries in crisis. Finally, Section 6 shows some conclusions and possible future work.

\section{Data and methodology}

In order to choose appropriate time series of the top stocks in terms of market capitalization belonging to the financial sector, we used the S\&P 1200 Global Index, which is a free-float weighted stock market index of stocks belonging to 31 countries. The stocks belonging to the index are responsible for approximately 70 percent of the total world stock market capitalization, and 200 of them belong to the financial sector, as classified by Bloomberg. From those, we extracted 197 stocks that had enough liquidity with respect to the working days of the New York Stock Exchange (NYSE). From the 197 stocks, 79 belong to the USA, 10 to Canada, 1 to Chile, 21 to the UK, 4 to France, 5 to Germany, 7 to Switzerland, 1 to Austria, 2 to the Netherlands, 2 to Belgium, 5 to Sweden, 1 to Denmark, 1 to Finland, 1 to Norway, 6 to Italy, 4 to Spain, 1 to Portugal, 1 to Greece, 12 to Japan, 9 to Hong Kong, 1 to South Korea, 1 to Taiwan, 3 to Singapore, and 18 to Australia. The stocks and their classification according to industry and sub-industry are listed in Appendix A.

We took the daily closing prices of each stock, and the resulting time series of all 197 stocks were compared with the time series of the NYSE, which was taken as a benchmark, since it is by far the major stock exchange in the world. If an element of the time series of a stock occurred for a day in which the NYSE wasn't opened, then this element was deleted from the time series, and if an element of the time series of a stock did not occur in a day in which the NYSE functioned, then we repeated the closing price of the previous day. The idea was not to eliminate too many days of the time series by, as an example, deleting all closing prices in a day one of the stock exchanges did not operate. The methodology which we chose would be particularly bad for stocks belonging to countries where weekends occur on different days than for Western countries, like Muslim countries or Israel, but since no stocks from our set belong to those countries, differences on weekends are not relevant here.

The data are organized so as to place stocks of the same country together, and then to discriminate stocks by industry and subindustry, according to the classification used by Bloomberg.  From the 197 stocks, 80 belong to Banks, 27 to Diversified Financial Services, 50 to Insurance Companies, 1 to an Investment Company, 1 to a Private Equity, 8 to Real Estate Companies, 28 are REITS (Real Estate Investment Trusts), and 2 belong to Savings \& Loans.

In order to reduce non-stationarity of the time series of the daily closing prices, we consider the log-returns of the closing prices, defined as
\begin{equation}
\label{logrets}
R_t=\ln (P_t)-\ln (P_{t-1})\ ,
\end{equation}
where $P_t$ is the closing price of the stock at day $t$ and $P_{t-1}$ is the closing price of the same stock at day $t-1$.

Since the stocks being considered belong to stock markets that do not operate at the same times, we run into the issue of lagging or not some stocks. Sandoval  (2012a), when dealing with stock market indices belonging to stock markets across the globe, showed that it is not very clear that an index has to be lagged with respect to another, except in cases like Japan and the USA. A solution is to use both original and lagged indices in the same framework, and to do all calculations as if the lagged indices were different ones. The same procedure is going to be followed here with the log-returns of the closing prices of the stocks that have been selected, so we shall deal with $2\times 197=394$ time series.

\section{Correlations}

Our first analysis of the data is based on the familiar correlation structure between the stocks. Correlation will be used in order to establish some methodology that will be followed later on for Transfer Entropy. The time series of log-returns of the 197 stocks with largest stock market capitalization and their lagged counterparts are used in order to calculate a correlation matrix $C$ whose elements $C_{ij}$ are the correlations between stocks $i$ and $j$. We use the usual Pearson correlation in this calculation, since previous results obtained with this type of (linear) correlation are in good accordance with results obtained using the Spearman rank correlation (Sandoval, 2013), which is more complex to calculate.

The structure of the resulting correlation matrix may be visualized in Figure 1a, where we plot a false color map of the elements of the correlation matrix, with lighter colors denoting higher correlations and darker ones denoting lower correlations. The figure displays the correlations in such a way that the leftmost and lowest corner corresponds to the correlation between element 1 with itself. The number of each stock grows from left to right and from the bottom to the top. The same configuration will be used in all other representations of matrices in this article. As expected, the diagonal elements are the brightest ones, with correlation 1 between all stocks and themselves. It is also possible to identify some clusters. First of all, there is a repetition pattern of stocks 1 to 197 and 198 to 394, corresponding to the original log-returns and the lagged ones in Figure 1a. If one plots the correlation matrix obtained by considering the original log-returns, plus lagged log-returns by one and by two days, the same structure repeats itself twice.

We may also identify other blocks, related with geographical position. The first one, going from 1 to 90, corresponds to stocks from North America (USA and Canada); the second one, from 91 to 152, corresponds to European stocks, plus Chile, which corresponds to a darker shade at 91 and which is here closer to Europe than to America; the third one is a loose structure of Australasian stocks, from 153 to 196. As said before, the pattern repeats itself for the lagged stocks. There are clear relations among these three main clusters, as shown by the brighter regions around the American and the European blocks, and also around the Australasian block. We may also see interaction of the Australasian block with the lagged stocks from America and Europe, showing a relation between Western stock markets with the Asian stock markets of the next day.

In Figure 1b, we highlighted the correlations between American and European stocks in the same day, and the correlations between European and Australasian stocks in the same day. Note that there are also weaker correlations between stocks of Australasia and stocks of America on the same day. In Figure 1c, we outline the correlations of the lagged stocks of America with the current stocks of Europe and of Australasia, and the correlations of lagged European stocks with the current stocks of Australasia. The figures show relations between stocks in the same day and between stocks of a previous day with stocks of the next day.

Within each main block, there are also some concentrations of brighter spots. This is more clearly visible if one plots only the elements of the correlation matrix that are above a certain correlation threshold. In Figure 2, we plot these values for the correlation matrix in black against a white background for threshold values 0.8, 0.7, 0.6, 0.5, 0.4, and 0.3, respectively. For 0.8, we find a structure of highly correlated stocks corresponding to REITS negotiated at the NYSE (original and lagged), between numbers 63 and 79. Between numbers 1 and 15, there is a loose agglomeration of Diversified Banking Institutions and Super-Regional-Banks of the USA. For threshold 0.7, there is a loose cluster of Banks from the USA (1 to 22), a tight cluster of Diversified Financial Services (Credit Card, Investment Management and Advisory Services) (27 to 33), a loose cluster of Insurance Companies (Multi-line, Life/Health, and Property/Casualty) (43 to 58), the cluster of REITS (63 to 79), a tight cluster of Canadian Commercial Banks (80 to 85), a small, but tight cluster of stocks from Life/Health Insurance Companies negotiated at the London Stock Exchange (102 to 105), another tight cluster of REITS negotiated at the UK (108 to 111), a tight cluster of stocks negotiated at the Paris Stock Exchange (112 to 115), immersed in a loose cluster of stocks negotiated in Germany, and of Banks from Switzerland (112 to 122), a tight cluster of stocks negotiated in Sweden (133 to 137), a cluster of stocks negotiated in Italy and Spain (141 to 150), a cluster of stocks negotiated in Japan (Banks and Diversified Financial Services) (154 to 160), a tight cluster of Real Estate companies from Japan (163 to 165), a cluster of stocks from Banks, Diversified Financial Services and Insurance from Hong Kong (167 to 172), a cluster made of a pair of stocks from Real Estate companies from Hong Kong (173 and 174), a cluster of Commercial Banks from Singapore (177 to 179), and a cluster of Commercial Banks from Australia. We also find some strong interactions between stocks from Italy and Spain with stocks from France, Germany, and Switzerland. The same pattern is followed by the lagged stocks.

For threshold 0.6, we already find some macro structures, like a cluster of stocks negotiated in the USA (1 to 79), interacting weakly with a cluster of stocks from Canada (80 to 89), a looser cluster of stocks from European countries (91 to 150), a cluster of stocks from Japan (153 to 165), a cluster of stocks from Hong Kong (166 to 174), a cluster of stocks from Singapore (177 to 179), and a cluster of Commercial Banks, one of Diversified Financial Services, and one of Insurance from Australia (180 to 184, and 186). For threshold 0.5, the macro structures according to continents become clearer for America and Europe, but Australasia is still fragmented. For 0.4, we start to see the correlations between the American and the European stocks, and an Australasian cluster becomes visible. We may also see a strong connection of an insurance company from Japan (number 162) with the lagged stocks of the USA and Canada. For threshold 0.3, the connections between America and Europe are strengthened, and we may also see connections of Australasia with European stocks of the same day, and with North American and European stocks of the previous day.

\subsection{Asset graphs}

Another way to analyze the structure of the correlations among the stocks of the financial sector here studied is to use asset graphs, which are based on a proper distance measure derived from the correlation matrix and on a threshold value for this distance, as in the works of Onnela, Chakraborti, Kaski, and Kertész (2002), Onnela, Chakraborti, and Kaski (2003), Onnela, Chakraborti, Kaski, and Kertész (2003a), Onnela, Chakraborti, Kaski, and Kertész (2003b),  Onnela, Kaski, and Kertész (2003), Sinha and Pan (2007), Ausloos and Lambiotte (2007), Sandoval (2012b), and Sandoval (2013). In an asset graph, given a certain threshold value, all distances below this threshold are represented as edges (links) between nodes, and all nodes without edges are not represented. This is a way of filtering some of the information and noise contained in a correlation matrix.

There are many ways to define a distance measure based on a correlation matrix, but the most used one in applications to financial markets is given by Mantegna (1999):
\begin{equation}
\label{distance}
d_{ij}=\sqrt{2\left( 1-c_{ij}\right) }\ ,
\end{equation}
where $c_{ij}$ is the correlation between nodes $i$ and $j$. As correlations between stocks vary from $-1$ (anticorrelated) to $1$ (completely correlated), the distance between them vary from $0$ (totally correlated) to $2$ (completely anticorrelated). Totally uncorrelated stocks would have distance $1$ between them.

Based on the distance measures, $m$-dimensional coordinates are assigned to each stock using an algorithm called Classical Multidimensional Scaling (Borg and Groener, 2005), which is based on minimizing the stress function
\begin{equation}
\label{stress}
S=\left[ \frac{\displaystyle{\sum_{i=1}^n\sum_{j>i}^n\left( \delta_{ij}-\bar d_{ij}\right) ^2}}{\displaystyle{\sum_{i=1}^n\sum_{j>i}^nd_{ij}^2}}\right] ^{1/2}\ \ ,\ \ \bar d_{ij}=\left[ \sum_{a=1}^m\left( x_{ia}-x_{ja}\right) ^2\right] ^{1/2}\ .
\end{equation}
where $\delta_{ij}$ is 1 for $i=j$ and zero otherwise, $n$ is the number of rows of the correlation matrix, and $\bar d_{ij}$ is an m-dimensional Euclidean distance (which may be another type of distance for other types of multidimensional scaling). The outputs of this optimization problem are the coordinates $x_{ia}$ of each of the nodes, where $i=1,\cdots, n$ is the number of nodes and $a=1,\cdots ,m$ is the number of dimensions in an $m$-dimensional space. The true distances are only perfectly representable in $m=n$ dimensions, but it is possible for a network to be well represented in smaller dimensions. In the case of this article we shall consider $m=2$ for a 2-dimensional visualization of the network, being the choice a compromise between fidelity to the original distances and the easiness of representing the networks in a two dimensional medium.

Figure 3a shows the stocks represented as nodes at the coordinates calculated by this procedure. White dots stand for the original log-returns, and black dots for their lagged values. There is a clear division between original and lagged stocks. Figure 3b represents the continents to which the stocks (original and lagged) belong, showing a clear division according to geography. The colors are black for America, white for Europe, and gray for Australasia. Note that the present stocks from Australasia are close both to the lagged stocks from America and to present stocks from Europe.

Figure 3 does not correspond to a network, since there are no edges between the nodes. By using the concept of asset graph, we may choose values for a distance threshold and represent only the edges that are below this threshold and the nodes connected by them. By choosing appropriate threshold values for the distance, above which edges and nodes are removed, we may obtain some filtered representations of the correlation structure between the stocks. Figure 4 presents the asset graphs for thresholds 0.6 and 0.8, in which we may see the formation of structures between the nodes. We did not represent the asset graphs for lower values than 0.6, because there are too few connections for them, nor higher values than 0.8, because the number of edges is so large that the pictures become hard to understand.

For threshold 0.4, the only connections are between the pairs Banco Bradesco and Itau Unibanco Holding (both stocks of Brazilian banks negotiated at the NYSE), Boston Properties (REITS - Office Porperty) and Vornado Realty Trust (REITS - Diversified), both from the USA, and Banco Bilbao Vizcaya Argentaria and Banco Santander (both commercial banks from Spain). For threshold 0.6, we have a large cluster comprised of REITS (Real Estate Investment Trusts), whose members are Apartment Investment \& Management, Avalon Bay Communities, Equity Residential (Apartments), Boston Properties (Office Property), Host Hotels \& Resorts (Hotels), Prologis (Warehouse/Industrial), Public Storage (Storage), Simon Property Group (Regional Malls), Kimco Realty (Shopping Centers), Ventas, HCP, Health Care REIT (Health Care),  Vornado Realty Trust, and Plum Creek Timber (Diversified). There is also a small cluster of banks from the USA, comprised of stocks of Bank of America, JP Morgan Chase, US Bancorp, and Wells Fargo, and the pairs The Goldman Sachs Group and Morgan Stanley	Banks (Diversified Banking Institutions), Comerica and BB\&T (Commercial Banks), The Bank of New York Mellon and Northern Trust (Fiduciary Banks), Franklin Resources and T Rowe Price Group (Investment Management / Advisory Services), Principal Financial Group and Prudential Financial (Life/Health Insurance), and The Travelers Cos and The Chubb (Property/Casualty Insurance). There is a small cluster of Canadian stocks, comprised of stocks of the Bank of Nova Scotia, the Royal Bank of Canada, and The Toronto-Dominion Bank, the pair of British REITS British Land and Land Securities Group, the small cluster of French banks, comprised of Crédit Agricole, BNP Paribas, and Société Générale, a small cluster of Japanes banks, Shinsei Bank, Sumitomo Mitsui Financial Group, and Mizuho Financial Group, the Japanese investment banks Daiwa Securities Group and Nomura Holdings, the Japanese real estate companies Mitsui Fudosan, Mitsubishi Estate, and Sumitomo Realty \& Development, the pair of banks from Hong Kong Industrial \& Commercial Bank of China and China Construction Bank Corp, and the pair of real estate companies from Hong Kong Cheung Kong Holdings and Sun Hung Kai Properties. The pairs Bradesco and Itau Unibanco, and Bilbao Vizcaya and Santander are still isolated from the other nodes.

For threshold 0.8, there is a large cluster of stocks of the USA, a cluster of stocks from Canada, a cluster of three banks of the UK, a cluster of four REITS from the UK, and a mixed cluster of European stocks. There are also four more clusters, one of Japanese stocks, another of Hong Kong stocks, a cluster of stocks of Singapore, and a cluster of stocks of Australia. For higher thresholds, the individual clusters merge more often, beginning with North America and Europe, and with the merging of the Australasian stocks, and then between the two main blocks and across original and lagged stocks. Figure 2 and the discussion associated with it is a good way to visualize the clustering that occurs here, since distance and correlation are related, although in a nonlinear way.

Some of the information obtained from the correlation matrix is plagued by noise, which may originate from the finiteness of data, residual non-stationarity of the time series, and many other sources. In order to gauge the effect of noise in the asset graphs, we calculated randomized time series for each stock, in which the order of the elements of each time series was randomly shuffled, so as to destroy any possible correlation between each time series but to preserve the frequency distribution of each one. We simulated 1,000 correlation matrices based on such shuffled time series, and calculated the distance matrix for each one. Excluding distances equal to zero, which is the distance between a stock and itself, we obtained a minimum distance equal to $d_{min}=1.265\pm 0.003$ (average $\pm$ standard deviation). It has been shown empirically by Sandoval (2013) that we may obtain more information about an asset graph if we consider thresholds that are close to this lower limit for noise. So, we shall consider the network of nodes whose edges are below the distance $d=1.2$.

\subsection{Centrality}

In network theory, the centrality of a node is important in the study of which nodes are, by some standard, more influential than others. Such measures may ne used, for instance, in the study of the propagation of epidemics, or the propagation of news, or, in the case of stocks, in the spreading of high volatility. There are various centrality measures (Newman, 2010), tending to different aspects of what we may think of ``central''. For undirected networks, for instance, we have Node Degree ($ND$), which is the total number of edges between a node and all others to which it is connected. This measure is better adapted to asset graphs, where not all nodes are connected between them, and varies according to the choice of threshold, as in Sandoval (2013). Another measure than can be used for asset graphs is Eigenvector Centrality ($EC$), which takes into account not just how many connections a node has, but also if it is localized in a region of highly connected nodes. There is also a measure called Closeness Centrality ($CC$) that measures the average distance (in terms of number of edges necessary to reach another node) of a certain node. This measure is larger for less central nodes, and if one wants a measure that, like the others, is larger for more central nodes, like the others we cited, then one may use Harmonic Closeness ($HC$), that is built on the same principles as Closeness Centrality, but is calculated using the inverse of the distances from one node to all others. The Betweenness Centrality ($BC$) of a node is another type of measure, that calculates how often a certain node is in the smaller paths between all other nodes. Still another measure of centrality, called Node Strength ($NS$), works for fully connected networks, and so is independent of thresholds in asset graphs, and takes into account the strength of the connections, which, in our case, are the correlations between the nodes. It measures the sum of the correlations of a node with all the others.

In Table 1, we present the nodes with highest centrality measures (top 5 values) by their names, countries they belong to, their industries and sub-industries. Names with an asterisk are lagged stocks. First, we find a preponderance of lagged stocks, except for Betweenness Centrality, since they are connected both among themselves and with the next days' values of stocks. There is also a preponderance of large banks (either listed as Diversified Banking Institutions or as Commercial Banks), with an important participation of Investment Management and Advisory Services for Node Degree and Eigenvector Centrality and a minor participation of Insurance companies. The USA dominates the scenario for Eigenvector centrality, since it has more stocks than the others, and they are more internally connected among themselves. Japan assumes preponderance in Betweenness centrality because of its role of connecting the lagged stocks of the USA and of Europe with the next day stocks of both continents. Node Strength has results that point mostly at European stocks, which have larger correlation values among themselves, in average.

\section{Transfer Entropy}

Although useful for determining which stocks behave similarly to others, the correlations between them cannot establish a relation of causality or of influence, since the action of a stock on another is not necessarily symmetric. A measure that has been used in a variety of fields, and which is both dynamic and non-symmetric, is {\sl Transfer Entropy}, developed by Schreiber (2000), which is based on the concept of {\sl Shannon Entropy}, first developed in the theory of information by Shannon (1948). Transfer entropy has been used in the study of cellular automata in Computer Science, in the study of the neural cortex of the brain, in the study of social networks, in Statistics, and also in the analysis of financial markets, as in the works of Kwon and Yang (2008a), Kwon and Yang (2008b), and Jizba, Kleinert, and Shefaat (2012), Baek, and Dimp, Huergo, and Peter (2012).

\scriptsize

\[ \begin{array}{c|l|l|l|l} \hline \text{\bf Centrality} & \text{\bf Company} & \text{\bf Country} & \text{\bf Industry} & \text{\bf Sub-Industry} \\ \hline \multicolumn{5}{c}{\text{\bf Node Degree}} \\ \hline 189 & \text{Credit Suisse Group*} & \text{France} & \text{Banks} & \text{Diversified Banking Institution} \\ 188 & \text{Franklin Resources*} & \text{USA} & \text{Diversif. Fin. Services} & \text{Investment Manag. / Adv. Services} \\ 187 & \text{Invesco*} & \text{USA} & \text{Diversif. Fin. Services} & \text{Investment Manag. / Adv. Services} \\ 185 & \text{T Rowe Price Group*} & \text{USA} & \text{Diversif. Fin. Services} & \text{Investment Manag. / Adv. Services} \\ 185 & \text{Zurich Insurance Group*} & \text{Switzerland} & \text{Insurance} & \text{Multi-line Insurance} \\ \hline \multicolumn{5}{c}{\text{\bf Eigenvector}} \\ \hline 0.089 & \text{Franklin Resources*} & \text{USA} & \text{Diversif. Fin. Services} & \text{Investment Manag. / Adv. Services} \\ 0.089 & \text{Invesco*} & \text{USA} & \text{Diversif. Fin. Services} & \text{Investment Manag. / Adv. Services} \\ 0.089 & \text{T Rowe Price Group*} & \text{USA} & \text{Diversif. Fin. Services} & \text{Investment Manag. / Adv. Services} \\ 0.089 & \text{Citigroup*} & \text{USA} & \text{Banks} & \text{Diversified Banking Institution} \\ 0.088 & \text{Itau Unibanco Holding*} & \text{USA} & \text{Banks} & \text{Commercial Bank} \\ 0.088 & \text{Prudential Financial*} & \text{USA} & \text{Insurance} & \text{Life/Health Insurance} \\ 0.088 & \text{Legg Mason*} &\text{USA} & \text{Diversif. Fin. Services} & \text{Investment Manag. / Adv. Services} \\ 0.088 & \text{The Goldman Sachs Group*} & \text{USA} & \text{Banks} & \text{Diversified Banking Institution} \\ 0.088 & \text{Ameriprise Financial*} & \text{USA} & \text{Diversif. Fin. Services} & \text{Investment Manag. / Adv. Services} \\ 0.088 & \text{Principal Financial Group*} & \text{USA} & \text{Insurance} & \text{Life/Health Insurance} \\ \hline \multicolumn{5}{c}{\text{\bf Harmonic Closeness}} \\ \hline 288 & \text{Franklin Resources*} & \text{USA} & \text{Diversif. Fin. Services} & \text{Investment Manag. / Adv. Services} \\ 286 & \text{Mitsubishi UFJ Financial Group} & \text{Japan} & \text{Banks} & \text{Diversified Banking Institution} \\ 283 & \text{Citigroup*} & \text{USA} & \text{Banks} & \text{Diversified Banking Institution} \\ 282.33 & \text{JP Morgan Chase \& Co*} & \text{USA} & \text{Banks} & \text{Diversified Banking Institution} \\ 282.17 & \text{Bank of America*} & \text{USA} & \text{Banks} & \text{Diversified Banking Institution} \\ \hline \multicolumn{5}{c}{\text{\bf Betweenness}} \\ \hline 3908 & \text{Mitsubishi UFJ Financial Group} & \text{Japan} & \text{Banks} & \text{Diversified Banking Institution} \\ 2272 & \text{Zurich Insurance Group} & \text{Switzerland} & \text{Insurance} & \text{Multi-line Insurance} \\  2126 & \text{Credit Suisse Group} & \text{France} & \text{Banks} & \text{Diversified Banking Institution} \\ 1970 & \text{National Australia Bank} & \text{Australia} & \text{Banks} & \text{Commercial Bank} \\  1819 & \text{Mitsubishi Estate Co} & \text{Japan} & \text{Real Estate} & \text{Real Estate Mang/Serv.}  \\ \hline \multicolumn{5}{c}{\text{\bf Node Strength}} \\ \hline 108.00 & \text{Deutsche Bank} & \text{Germany} & \text{Banks} & \text{Diversified Banking Institution} \\ 107.85 & \text{Franklin Resources*} & \text{USA} & \text{Diversif. Fin. Services} & \text{Investment Manag. / Adv. Services} \\ 106.52 & \text{Zurich Insurance Group} & \text{Switzerland} & \text{Insurance} & \text{Multi-line Insurance} \\ 105.20 & \text{Credit Suisse Group} & \text{France} & \text{Banks} & \text{Diversified Banking Institution} \\ 105.09 & \text{Allianz} & \text{Germany} & \text{Insurance} & \text{Multi-line Insurance} \\ \hline \end{array} \]

\normalsize

\noindent {\bf Table 1.} Classification of stocks with highest centrality measures, the countries they belong to, their industry and sub-industry classifications, for threshold 1.2. Only the five stocks with highest centrality values are shown (more, in case of draws). The names with an asterisk are lagged stocks.

\vskip 0.3 cm

In this section, we shall describe the concept of Transfer Entropy (TE), using it to analyze the data concerning the 197 stocks of companies of the financial sector and their lagged counterparts. We will start by describing briefly the concept of Shannon entropy.

\subsection{Shannon Entropy}

The American mathematician, electronic engineer and cryptographer, Claude Elwood Shannon (1916–2001), founded the theory of information in his work ``A Mathematical Theory of Communication'' (Shannon, 1948), in which he derived what is now known as the {\sl Shannon entropy}. According to Shannon, the main problem of information theory is how to reproduce at one point a message sent from another point. If one considers a set of possible events whose probabilities of occurrence are $p_i$, $i=1,\cdots ,n$, then a measure $H(p_1,p_2,\cdots ,p_n)$ of the uncertainty of the outcome of an event given such distribution of probabilities should have the following three properties:

\vskip 0.1 cm

\noindent $\bullet $ $H(p_i)$ should be continuous in $p_i$;

\noindent $\bullet $ if all probabilities are equal, what means that $p_i=1/n$, then $H$ should be a monotonically increasing function of $n$ (if there are more choices of events, then the uncertainty about one outcome should increase);

\noindent $\bullet $ if a choice is broken down into other choices, with probabilities $c_j$, $j=1,\cdots ,k$, then $H=\sum_{j=1}^kc_jH_k$, where $H_k$ is the value of the function $H$ for each choice.

\vskip 0.1 cm

Shannon proved that the only function that satisfies all three properties is given by
\begin{equation}
\label{Shannon}
H=-\sum_{i=1}^Np_i\log_2p_i\ ,
\end{equation}
where the sum is over all states for which $p_i\neq 0$ (Shannon's definition had a constant $k$ multiplied by it, which has been removed here). The base 2 for the logarithm is chosen so that the measure is given in terms of bits of information. As an example, a device with two positions (like a flip-flop circuit) can store one bit of information. The number of possible states for $N$ such devices would then be $2^N$, and $\log_22^N=N$, meaning that $N$ such devices can store $N$ bits of information, as should be expected. This definition bears a lot of resemblance to Gibbs' entropy, but is more general, as it can be applied to any system that carries information.

The Shannon entropy represents the average uncertainty about measures $i$ of a variable $X$ (in bits), and quantifies the average number of bits needed to encode the variable $X$. In the present work, given the time series of the log-returns of a stock, ranging over a certain interval of values, one may divide such possible values into $N$ different bins and then calculate the probabilities of each state $i$, what is the number of values of $X$ that fall into bin $i$ divided by the total number of values of $X$ in the time series. The Shannon entropy thus calculated will depend on the number of bins that are selected. After selecting the number of bins, one associates a symbol (generally a number) to each bin.

Using the stocks of the J.P. Morgan (code JPM), classified as a Diversified Banking Institution, we shall give an example of the calculation of the Shannon Entropy for two different choices of bins. In Figure 5, we show the frequency distributions of the log-returns for the stocks of the J.P. Morgan from 2007 to 2012, which varied from -0.2323 to 0.2239 during that period, with two different binning choices. The first choice results in 24 bins of size 0.02, and the second choice results in 6 bins of size 0.1.

To each bin is assigned a symbol, which, in our case, is a number, from 1 to 24 in the first case and from 1 to 6 in the second case. Figure 6 shows the assigning of symbols for the two choices of binning for the first log-returns of the stocks of the J.P. Morgan. Then, we calculate the probability that a symbol appears in the time series and then use (\ref{Shannon}) in order to calculate the Shannon entropy, which, in our case, is $H=2.55$ for bins of size 0.02 and $H=0.59$ for bins of size 0.1. The second result is smaller than the first one because there is less information for the second choice of binning due to the smaller number of possible states of the system. The difference in values, though, is not important, since we shall use the Shannon entropy as a means of comparing the amount of information in different time series.

Figure 7 shows the Shannon Entropy calculated for each stock in this study (the lagged stocks are not represented, since their entropies are nearly the same as the entropies of the original stocks). The results for both choices of binning are in fact very similar, and their correlation is 0.97. Stocks with higher Shannon Entropy are the most volatile ones. As one can see, the second choice, with larger bin sizes, shows the differences more sharply, which is one of the reasons why larger binnings are usually favored in the literature.

\subsection{Transfer Entropy}

When one deals with variables that interact with one another, then the time series of one variable $Y$ may influence the time series of another variable $X$ in a future time. We may assume that the time series of $X$ is a Markov process of degree $k$, what means that a state $i_{n+1}$ of $X$ depends on the $k$ previous states of the same variable. This may be made more mathematically rigorous by defining that the time series of $X$ is a Markov state of degree $k$ if
\begin{equation}
\label{Markov}
p\left( i_{n+1}|i_n,i_{n-1},\cdots ,i_0\right) =p\left( i_{n+1}|i_n,i_{n-1},\cdots ,i_{n-k+1}\right) \ ,
\end{equation}
where $p(A|B)$ is the conditional probability of $A$ given $B$, defined as
\begin{equation}
\label{conditional}
p(A|B)=\frac{p(A,B)}{p(B)}\ .
\end{equation}
What expression (\ref{Markov}) means is that the conditional probability of state $i_{n+1}$ of variable $X$ on all its previous states is the same as the conditional probability of $i_{n+1}$ on its $k$ previous states, meaning that it does not depend on states previous to the $k$th previous states of the same variable.

One may also assume that state $i_{n+1}$ of variable $X$ depends on the $\ell $ previous states of variable $Y$. The concept is represented in Figure 8, where the time series of a variable $X$, with states $i_n$, and the time series of a variable $Y$, with states $j_n$, are identified.

We may now define the concept of Transfer Entropy from a time series $Y$ to a times series $X$ as the average information contained in the source $Y$ about the next state of the destination $X$ that was not already contained in the destination's past. We assume that element $i_{n+1}$ of the time series of variable $X$ is influenced by the $k$ previous states of the same variable and by the $\ell $ previous states of variable $Y$. The values of $k$ and $\ell $ may vary, according to the data that is being used, and to the way one wishes to analyze the transfer of entropy of one variable to the other.

Transfer Entropy from variable $Y$ to variable $X$ is defined as
\begin{eqnarray}
\label{transferentropy}
TE_{Y\rightarrow X}(k,\ell ) & = & \sum_{i_{n+1},i_n^{(k)},j_n^{(\ell )}}p\left( i_{n+1},i_n^{(k)},j_n^{(\ell )}\right) \log_2p\left( i_{n+1}|i_n^{(k)},j_n^{(\ell )}\right) \nonumber \\ & & -\sum_{i_{n+1},i_n^{(k)},j_n^{(\ell )}}p\left( i_{n+1},i_n^{(k)},j_n^{(\ell )}\right) \log_2p\left( i_{n+1}|i_n^{(k)}\right) \nonumber \\
& = & \sum_{i_{n+1},i_n^{(k)},j_n^{(\ell )}}p\left( i_{n+1},i_n^{(k)},j_n^{(\ell )}\right) \log_2\frac{p\left( i_{n+1}|i_n^{(k)},j_n^{(\ell )}\right) }{p\left( i_{n+1}|i_n^{(k)}\right) }\ ,
\end{eqnarray}
where $i_n$ is element $n$ of the time series of variable $X$ and $j_n$ is element $n$ of the time series of variable $Y$, $p(A,B)$ is the joint probability of $A$ and $B$, and
\begin{equation}
p\left( i_{n+1},i_n^{(k)},j_n^{(\ell )}\right) =p\left( i_{n+1},i_n,\cdots ,i_{n-k+1},j_n,\cdots ,j_{n-\ell +1}\right) \end{equation}
is the joint probability distribution of state $i_{n+1}$, of state $i_n$ and its $k$ predecessors, and the $\ell $ predecessors of state $j_n$, as in Figure 8.

This definition of Transfer Entropy assumes that events on a certain day may be influenced by events of $k$ and $\ell $ previous days. We shall assume, with some backing from empirical data for financial markets, that only the day before is important. By doing so, formula (\ref{transferentropy}) for the Transfer Entropy of $Y$ to $X$ becomes simpler:
\begin{equation}
\label{TE}
TE_{Y\rightarrow X}=\sum_{i_{n+1},i_n,j_n}p\left( i_{n+1},i_n,j_n\right) \log_2\frac{p\left( i_{n+1}|i_n,j_n\right) }{p\left( i_{n+1}|i_n\right) }=\sum_{i_{n+1},i_n,j_n}p\left( i_{n+1},i_n,j_n\right) \log_2\frac{p\left( i_{n+1},i_n,j_n\right) p\left( i_n\right) }{p\left( i_{n+1},i_n\right) p\left( i_n,j_n\right) }\ ,
\end{equation}
where we took $k=\ell =1$, meaning we are using lagged time series of one day, only.

In order to exemplify the calculation of Transfer Entropy, we will now show some steps for the calculation of the Transfer Entropy from the Deutsche Bank to the J.P. Morgan. In Figure 9, first table, we show the initial part of the time series for the log-returns of the J.P. Morgan, which we call vector $X_{n+1}$ (first column), for its values lagged by one day, vector $X_n$ (second column), and the log-returns of the Deutsche Bank lagged by one day, vector $Y_n$ (third column). Calculating the minimum and maximum returns of the entire set of time series, we obtain a minimum value $m=-1.4949$ and a maximum value $M=0.7049$. Considering then an interval $[-1.5,0.8]$ with increments $0.1$, we obtain 24 bins to which we assign numeric symbols going from 1 to 24. Then, we associate one symbol to each log-return, depending on the bin it belongs to. As seen in Figure 9, second table, most of the symbols orbit around the intervals closest to zero, since most of the variations of the time series are relatively small.

In order to calculate the simplest probabilities, $p(i_n)$ appearing in (\ref{TE}), we just need to count how many times each symbol appears in vector $X_n$ and then divide by the total number of occurrences. As an example, from the first 10 lines of data shown in Figure 9, the symbol 15 appears 4 times. In order to calculate $p\left( i_{n+1},i_n\right) $, we must count how many times a particular combination of symbols, $(a,b)$, appears in the joint columns $X_{n+1}$ and $X_n$. As an example, in the first ten lines of such columns, the combination $(15,15)$ appears zero times, the combination $(15,16)$ appears 4 times, the combination $(16,15)$ appears 4 times, and the combination $(16,16)$ appears two times.

We now sum over all combinations of the components of $X_{n+1}$, $X_n$, and $Y_n$ using definition (\ref{TE}), obtaining as a result $TE_{177\rightarrow 4}=0.0155$. This result indicates the average amount of information transferred from the Deustche Bank to the J.P. Morgan which was not already contained in the information of the past state of the J.P. Morgan one day before. Doing the same for all possible combinations of stocks, one obtains a Transfer Entropy matrix, which is represented in terms of false colors in Figure 11a.

Here, like in the calculation of the Shannon Entropy, the size of the bins used in the calculations of the probabilities changes the resulting Transfer Entropy (TE). The calculations we have shown in figures 9 and 10 are relative to a choice of binning of size 0.1. In order to compare the resulting TE matrix with that of another choice for binning, we calculated the TE for binning size 0.02, what leads to a much larger number of bins and to a much longer calculation time. The resulting TE matrix for binning 0.02 is plotted in Figure 11b. The two TE matrices are not very different, with the main dissimilarities being due to scale. The visualization for binning size $0.1$ is sharper than the one obtained using binning size $0.02$. In what follows, we shall consider binning size 0.1 throughout the calculations, since it demands less computation time and delivers clearer results in comparison with the ones obtained for some smaller sized binnings.

\subsection{Effective Transfer Entropy}

Transfer Entropy matrices usually contain much noise, due to the finite size of data used in their calculation, non-stationarity of data, and other possible effects, and we must also consider that stocks that have more entropy, what is associated with higher volatility, naturally transfer more entropy to the others. We may eliminate some of these effects if we calculate the Transfer Entropy of randomized time series, where the elements of each time series are randomly shuffled so as to break any causality relation between variables but maintain the individual probability distributions of each time series. The original Transfer Entropy matrix is represented in Figure 12a. The result of the average of 25 simulations with randomized data appears in Figure 12b. We only calculated 25 simulations because the calculations are very computationally demanding, and because the results for each simulation are very similar. Then, an Effective Transfer Entropy matrix (ETE) may be calculated by subtracting the Randomized Transfer Entropy matrix (RTE) from the Transfer Entropy matrix (TE):
\begin{equation}
\label{ETE}
ETE_{Y\rightarrow X}=TE_{Y\rightarrow X}-RTE_{Y\rightarrow X}\ .
\end{equation}
The result is shown in Figure 12c.

The main feature of the representation of the Effective Transfer Entropy matrix (or of the Transfer Entropy matrix) is that it is clearly not symmetric. The second one is that the highest results are all in the quadrant on the left topmost corner (Quadrant 12). That is the quadrant related with the Effective Transfer Entropy (ETE) from the lagged stocks to the original ones. The main diagonal expresses the ETE from one stock to itself on the next day, which, by the very construction of the measure being used, is expected to be high. But Quadrant 12 also shows that there are larger transfers of entropy from lagged stocks to the other ones than between stocks on the same day. We must remind ourselves that we are dealing here with the daily closing prices of stocks, and that the interaction of prices of stocks, and their reactions to news, usually occur at high frequency. Here, we watch the effects that a whole day of negotiation of a stock has on the others. Figure 13a shows a closer look at the ETE of the stocks on stocks on the same day, what corresponds to the quadrant on the bottom left (Quadrant 11), and from lagged to original stocks, in Figure 13b (Quadrant 12).

Analyzing Quadrant 12 (Figure 13b), we may see again the structures due to geographical positions, with clusters related with stocks from the USA (1 to 79), Canada (80 to 89), Europe (91 to 152), Japan (153 to 165), Hong Kong (166 to 174), Singapore (177 to 179), and Australia (180 to 197). We also detect some ETE from lagged stocks from the USA to stocks from Canada and Europe, from lagged stocks from Europe to stocks from the USA and Canada and, with a smaller strength, from lagged stocks from Europe to stocks from Australasia, and transfer of entropy within the Australasian stocks. Quadrant 11 (Figure 13a) shows much smaller values, but one can see a clear influence of Japan (153-165) on North America (1-89) and Europe (91-152), and also some influence from Europe to the USA. A very light influence may be seen from the USA to itself on the next day, Canada, and Europe, but it is already hard to distinguish this influence from noise. There are negative values of ETE, what means that the Transfer Entropy calculated is smaller than what would be expected from noise.

There are intra-sector structures inside each block, but this may be best analyzed by using thresholds above which we assign value 1 to ETEs, and bellow which we assign value 0. Figure 14 shows the resulting false color maps for thresholds 0.3, 0.2, and 0.1. The structures previously described are all quite clear in these graphs for Quadrant 12. We shall discuss in more detail the structure that appears from threshold 0.4, which is not shown in Figure 14, since it has very few connections, with the main ones being the ETEs from lagged stocks to their original counterparts. At this threshold, for stocks of the USA, there is already an ETE from the State Street (Fiduciary Bank) to the Fifth Third Bancorp (Super-regional Bank), and mutual exchanges of ETE between Prudential Financial (Life/Health Insurance) and MetLife (Multi-line Insurance), between Itau Unibanco Holding and Banco Bradesco (both stocks of two Brazilian banks negotiated in the NYSE), and between HCP and Ventas (both REITS-Health Care). There is also a dense cluster of REITS, with ETEs flowing from one to the other, but not from all of them to all of them, composed of Apartment Investment \& Management and Equity Residential (REITS-Apartments), Boston Properties (REITS-Office Property), Simon Property Group (REITS-Regional Malls), Kimco Realty (REITS-Shopping Centers), and Vornado Realty Trust (REITS-Diversified). The most pointed to stock is the one of Vornado Realty Trust. From Spain, we have a mutual relation between Banco Bilbao Vizcaya Argentaria and Banco Santander (both large Commercial Banks). From Japan, there is a pair of interdependent stocks, Mitsubishi UFJ Financial (Diversified Banking Institution) and Sumitomo Mitsui Financial (Commercial Banks), and a trio, consisting of Mitsui Fudosan and Sumitomo Realty \& Development (both Real Estate Operation/Development), and Mitsubishi Estate (Real Estate Management/Services). A last pair occurs for Hong Kong, between the stocks of China Construction Bank and Industrial \& Commercial Bank of China (both Commercial Banks).

\subsection{Normalized Transfer Entropy and Asset Graphs}

We may again try to produce a map of the nodes according to distances between stocks. The problem now is that distance is a symmetric measure, and the Effective Transfer Entropy is not. Another problem is that the ETE is not normalized. We may correct the latter problem by defining the {\sl Normalized Transfer Entropy}, which uses another measure derived from the Shannon entropy, called {\sl Conditional Entropy}, which is defined in the following way: the Conditional Entropy of $X$ given $Y$ is the average uncertainty in the outcome of a measurement $x$ of $X$ when the measure $y$ of $Y$ is known:
\begin{equation}
\label{conditionalentropy}
H_{X|Y}=-\sum_{i_n,j_n}p\left( i_n,j_n\right) \log_2p\left( i_n|j_n\right) =-\sum_{i_n,j_n}p\left( i_n,j_n\right) \log_2\frac{p\left( i_n,j_n\right) }{p\left( j_n\right) }\ .
\end{equation}

Based on this concept, we may define the {\sl Normalized Transfer Entropy} as
\begin{equation}
\label{NTE}
NTE=\frac{ETE_{Y\rightarrow X}}{H_{X^F|X^P}}\ ,
\end{equation}
where $H_{X^F|X^P}$ is the conditional entropy of the future of $X$ on its past, what we may write as
\begin{equation}
H_{X^F|X^P}=-\sum_{i_n,j_n}p\left( i_{n+1},i_n\right) \log_2\frac{p\left( i_{n+1},i_n\right) }{p\left( i_n\right) }\ .
\end{equation}

The resulting values are always between -1 and 1. Using now definition (\ref{distance}), we may define elements $d_{ij}$. However, the resulting matrix does not necessarily have $d_{ii}=0$, what is a necessary condition for it to be a distance measure. So we must fix that by setting all diagonal elements to zero. The resulting matrix is still not symmetric, and we symmetrize the matrix by setting $d_{ij}=d_{ji}$ if $d_{ij}>d_{ji}$ and $d_{ji}=d_{ij}$, otherwise, what means that we always consider the smallest between the two values $d_{ij}$ and $d_{ji}$ to be the distance between $i$ and $j$. The resulting distance matrix is then used, applying (\ref{stress}), in order to calculate a set of coordinates for each stock as a node in a space where distances are similar to the ones given by the symmetrized distance matrix.

Figure 15 shows the stocks (original and lagged ones) plotted in two-dimensional graphs. In Figure 15a, original stocks are in white and their lagged values are in black. As expected from the results we saw for the ETE matrix, lagged and original values are very close one to the other. This is in strong contrast with the results obtained using correlation (Figure 3) where original and lagged stocks occupy very distinct positions. In Figure 15b, the lagged stocks were removed, and continents are highlighted with different shades of gray: white for America, black for Europe and gray for Australasia. Another difference that may be seen here is that Australasia seems closer to America than Europe.

Once more, by using thresholds, we are able to filter some of the information in such graph, and we may also build asset graphs with connections between some nodes. Here, we choose the values of the ETE and not of the distance matrix in order to establish thresholds. The first reason is because the distance matrix highly modifies the original relations between stocks and lagged stocks, and the second one is that the distance values do not vary very linearly. For a choice of thresholds 0.4, 0.3, and 0.2 for the ETE, deleting all edges below these values and all unconnected nodes after that removal, we obtain the graphs in Figure 16. The number of connections (edges) increases dramatically for higher values of the threshold, approaching a limit at which all nodes are connected.

In Figure 17, we take a closer look at the relationships between the stocks at threshold 0.4. At the lower right corner, there are three small clusters of stocks from the USA in the same rectangle. The first one is the transfer entropy between stocks of Well Fargo (Super-Regional Bank) to the stocks of J.P. Morgan Chase (Diversified Banking Institution); the second one is a cluster of Insurance companies (Hartford, Principal, Met Life, Prudential, and Lincoln); the third one is a small cluster of Super-Regional Banks (Huntington Bancshares, Fifth Third, and Sun Trust). At the top right rectangle, there are two clusters of stocks from the USA. The first one is a large cluster of REITS (Real Estate Investment Trusts), comprising Avalon Bay, Equity Residential, Apartment Investment \& Management, Kimko Realty, Macerich, Simon Property Group, Boston Properties, Prologis, and Vornado Realty Trust; the second one is a pair of two REITS of Health Care: HCP and Ventas. At the center of the graph, we have a rectangle with the pair Banco Bradesco and Itau Unibanco, which are the stocks of major Commercial Banks based in Brazil negotiated in the New York Stock Exchange. At the lower left of the graph, there are two pairs: one of Diversified Banking Institutions from France (Societé Générale and BNP Paribas) and one of major Commercial Banks from Spain (Banco Bilbao Vizcaya Argentaria and Banco Santander). At the top left, we have the last clusters; the first one, a pair of stocks from Japan: Mitsubishi UFJ Financial Group (Diversified Banking Institution) and of Sumimoto Mitsui Financial Group (Commercial Bank); the second one, whose elements are Mitsubishi Estate, Mitsui Fudosan, and Sumitomo Realty \& Development, is a cluster of Real Estate operations, management and services firms; the third one is a pair of two Commercial Banks from Hong Kong: Industrial \& Commercial Bank of China and China Construction Bank. It is to be noticed that most relations are reciprocate, although the ETE between stocks is rarely very similar.

We shall not make a deeper analysis of the remaining asset graphs, but one can see that integration begins inside countries, with the exception of certain countries from Europe, and then goes continental. Only at threshold 0.1 and below, we start having intercontinental integration. This may be due to differences in operation hours of the stock exchanges, to geographical, economic and cultural relations, or to other factors we failed to contemplate (see, for instance, Sandoval, 2012a for a discussion).

\subsection{Centralities}

The measures of centrality presented in Section 3 are appropriate for an undirected network, like the one obtained by using correlation, but the networks built using Effective Transfer Entropy are directed nodes, that have either ingoing edges to a node, outgoing edges from the node, or both. So, centrality measures often break down into ingoing and outgoing ones. As an example, a node may be highly central with respect to pointing at other nodes, like the Google search page; these are called {\sl hubs}. Other nodes may have many other nodes pointing at it, as in the case of a highly cited article in a network of citations; these are called {\sl authorities}. Each one is central in a different way, and a node may be central according to both criteria. Node degree, for example, may be broken in two measures: In Node Degree ($ND_{in}$), which measures the sum of all ingoing edges to a certain node, and Out Node Degree ($ND_{out}$), which measures the sum of all outgoing edges from a node. In a similar way, one defines In Eigenvector Centrality ($EC_{in}$) and Out Eigenvector Centrality ($EC_{in}$), and In Harmonic Closeness ($HC_{in}$) and Out Harmonic Closeness ($HC_{in}$). Betweenness Centrality is now calculated along directed paths only, and it is called Directed Betweenness Centrality, $(BC_{dir}$).

As we said before, when applying centrality measures to asset graphs, those measures vary according to the chosen value for the threshold. As extreme examples, if the threshold is such that the network has very few nodes, Node Centrality, for example, will also be low. If the threshold value is such that every node is connected to every other node, then all Node Degrees will be the same: the number of all connections made between the nodes. It has been shown empirically (Sandoval, 2013) that one gets the most information about a set of nodes if one considers asset graphs whose thresholds are close to the minimum or the maximum of the values obtained through simulations with randomized data. We may rephrase it by saying that we obtain more information of a network when we consider its limit to results obtained from noise. From the simulations we have made in order to calculate the Effective Transfer Entropy, we could check that the largest values of Transfer Entropy for randomized data are close to 0.05 for the choice of bins with size 0.1 (Figure 12a). So, we shall consider here the centrality measures that were mentioned applied to the directed networks obtained from the Effective Transfer Entropy with threshold 0.05. The results are plotted in Figure 18. As the values of different centralities may vary a lot (from 3 to 153 for $ND_{in}$ and from 0 to 1317 for $BC_{dir}$), we normalize all centrality measures by setting their maxima to one. For all but Directed Betweenness Centrality, stocks belonging to the Americas and to Europe appear more central.

Table 2 presents the most central stocks according to each centrality measure. Only the first five stocks are shown (more, in case of draws). Lagged stocks appear with an $*$ besides the names of the companies. Since we are considering only the strong values of transfer entropy, and since asset graphs do not involve the nodes that are not connected, this excludes all connections, except the ones between lagged and original log-returns. So, all in degrees are of original stocks and all out degrees (including Directed Betweenness) are of lagged stocks. For out degrees, insurance companies occupy the top positions, together with some banks, all of them belonging to European or to U.S. companies. For in degrees, we see a predominance of banks, but insurance companies also occupy top positions. This means there is a tendency of entropy being transferred from insurance companies to banks. For Directed Betweeenness, the top positions are occupied by major European banks and also by other types of companies.

Figure 19 shows the normalized values of the centrality measures for the asset graph obtained with threshold 0.1. The figure has a smaller number of stocks, since there are slightly fewer nodes for this value of the threshold. One may notice a sharp drop in values for Eigenvector Centralities in this asset graph. Table 3 shows the most central stocks according to each centrality measure for this choice of binning. Only the first five stocks are shown (more, in case of draws). In all centrality measures, insurance companies occupy the first positions, and the same stocks usually occupy these positions, except for Provident Financial.

For threshold 0.2, there is also a preponderance on insurance companies and banks from the USA, and for thresholds 0.3 and 0.4, there are mostly banks and REITS occupying the first positions, also due to the fact that they are some of the only nodes that are part of the asset graphs at these threshold values.

The centrality measures we have considered thus far in this section do not take into account the strength of the connections between the nodes. There are centrality measures that take that into account, being the main one called {\sl Node Strength} ($NS$), which, in undirected networks, is the sum of all connections made by a node. For directed networks, we have the {\sl In Node Strength} ($NS_{in}$), which measures the sum of all ingoing connections to a node,  and the {\sl Out Node Strength} ($NS_{out}$), which measures the sum of all outgoing connections from a node. These are centrality measures that can be applied to the whole network, including all nodes. Figure 20 shows the results for both centrality measures, and Table 4 shows the top five stocks according to each node centrality. We used ETE in the calculations. Had we used TE instead, the results would be the same.

The five top stocks for In Node Strength are those of Insurance Companies, qualified as authorities, which are nodes to which many other nodes point, and with high values of ETE, what means that there is a large amount of information flowing into the log-returns of those stocks. For Out Node Strength, again insurance companies dominate, what means that they send much information into the prices of the other stocks (they are also hubs).

\scriptsize

\[ \begin{array}{c|l|l|l|l} \hline \text{\bf Centrality} & \text{\bf Company} & \text{\bf Country} & \text{\bf Industry} & \text{\bf Sub-Industry} \\ \hline \multicolumn{5}{c}{\text{\bf In Node Degree}} \\ \hline 153 & \text{Credit Suisse Group AG} & \text{Switzerland} & \text{Banks} & \text{Diversified Banking Inst}\\ 150 & \text{Deutsche Bank AG} & \text{Germany} & \text{Banks} & \text{Diversified Banking Inst} \\ 149 & \text{Invesco} & \text{USA} & \text{Diversified Finan Serv} & \text{Invest Mgmnt/Advis Serv} \\ 149 & \text{ING Groep NV} & \text{Netherlands} & \text{Insurance} & \text{Life/Health Insurance} \\ 149 & \text{KBC Groep NV} & \text{Belgium} & \text{Banks} & \text{Commer Banks Non-US}\\ \hline \multicolumn{5}{c}{\text{\bf Out Node Degree}} \\ \hline 160 & \text{ING Groep NV*} & \text{Netherlands} & \text{Insurance} & \text{Life/Health Insurance} \\ 158 & \text{Hartford Financial Services Group*} & \text{USA} & \text{Insurance} & \text{Multi-line Insurance} \\ 154 & \text{KBC Groep*} & \text{Belgium} & \text{Banks} & \text{Commer Banks Non-US} \\ 152 & \text{Genworth Financial*} & \text{USA} & \text{Insurance} & \text{Multi-line Insurance} \\ 151 & \text{Lincoln National Corp*} & \text{USA} & \text{Insurance} & \text{Life/Health Insurance}\\ \hline \multicolumn{5}{c}{\text{\bf In Eigenvector}} \\ \hline 11.99 & \text{Invesco} & \text{USA} & \text{Diversified Finan Serv} & \text{Invest Mgmnt/Advis Serv} \\ 11.91 & \text{Credit Suisse Group AG} & \text{Switzerland} & \text{Banks} & \text{Diversified Banking Inst} \\ 11.86 & \text{Hartford Financial Services Group} & \text{USA} & \text{Insurance} & \text{Multi-line Insurance} \\ 11.85 & \text{Lincoln National Corp} & \text{USA} & \text{Insurance} & \text{Life/Health Insurance} \\ 11.83 & \text{MetLife} & \text{USA} & \text{Insurance} & \text{Multi-line Insurance}\\ \hline \multicolumn{5}{c}{\text{\bf Out Eigenvector}} \\ \hline 0.094 & \text{Hartford Financial Services Group*} & \text{USA} & \text{Insurance} & \text{Multi-line Insurance} \\ 0.094 & \text{Lincoln National Corp*} & \text{USA} & \text{Insurance} & \text{Life/Health Insurance} \\ 0.093 & \text{Invesco*} & \text{USA} & \text{Diversified Finan Serv} & \text{Invest Mgmnt/Advis Serv} \\ 0.093 & \text{MetLife*} & \text{USA} & \text{Insurance} & \text{Multi-line Insurance} \\ 0.093 & \text{ING Groep*} & \text{Netherlands} & \text{Insurance} & \text{Life/Health Insurance} \\ 0.093 & \text{Genworth Financial*} & \text{USA} & \text{Insurance} & \text{Multi-line Insurance} \\ 0.093 & \text{Principal Financial Group*} & \text{USA} & \text{Insurance} & \text{Life/Health Insurance} \\ 0.093 & \text{UBS*} & \text{Switzerland} & \text{Banks} & \text{Diversified Banking Inst} \\ 0.093 & \text{Prudential Financial*} & \text{USA} & \text{Insurance} & \text{Life/Health Insurance} \\ 0.093 & \text{Ameriprise Financial*} & \text{USA} & \text{Diversified Finan Serv} & \text{Invest Mgmnt/Advis Serv}\\ \hline \multicolumn{5}{c}{\text{\bf In Harmonic Closeness}} \\ \hline 174.00 & \text{Credit Suisse Group AG} & \text{Switzerland} & \text{Banks} & \text{Diversified Banking Inst} \\ 172.5 & \text{Deutsche Bank AG} & \text{Germany} & \text{Banks} & \text{Diversified Banking Inst} \\ 171.8 & \text{KBC Groep NV} & \text{Belgium} & \text{Banks} & \text{Commer Banks Non-US} \\ 171.2 & \text{ING Groep NV} & \text{Netherlands} & \text{Insurance} & \text{Life/Health Insurance} \\ 170.5 & \text{Commerzbank AG} & \text{Germany} & \text{Banks} & \text{Commer Banks Non-US} \\ \hline \multicolumn{5}{c}{\text{\bf Out Harmonic Closeness}} \\ \hline 178 & \text{ING Groep*} & \text{Netherlands} & \text{Insurance} & \text{Life/Health Insurance} \\ 177 & \text{Hartford Financial Services Group*} & \text{USA} & \text{Insurance} & \text{Multi-line Insurance} \\ 175 & \text{KBC Groep*} & \text{Belgium} & \text{Banks} & \text{Commer Banks Non-US} \\ 174 & \text{Genworth Financial*} & \text{USA} & \text{Insurance} & \text{Multi-line Insurance} \\ 173 & \text{Barclays*} & \text{UK} & \text{Banks} & \text{Diversified Banking Inst} \\ \hline \multicolumn{5}{c}{\text{\bf Directed Betweenness}} \\ \hline 1317 & \text{KBC Groep*} & \text{Belgium} & \text{Banks} & \text{Commer Banks Non-US} \\ 1202 & \text{China Construction Bank Corp*} & \text{Hong Kong} & \text{Banks} & \text{Commer Banks Non-US} \\ 1074 & \text{ING Groep*} & \text{Netherlands} & \text{Insurance} & \text{Life/Health Insurance} \\ 998 & \text{Goodman Group*} & \text{Australia} & \text{REITS} & \text{REITS-Diversified}\\ 984 & \text{Barclays*} & \text{UK} & \text{Banks} & \text{Diversified Banking Inst}\\ \hline \end{array} \]

\normalsize

\noindent {\bf Table 2.} Classification of stocks with highest centrality measures, the countries they belong to, their industry and sub-industry classifications, for asset graphs based on threshold 0.05. Only the five stocks with highest centrality values are shown (more, in case of draws).

\vskip 0.3 cm

\scriptsize

\[ \begin{array}{c|l|l|l|l} \hline \text{\bf Centrality} & \text{\bf Company} & \text{\bf Country} & \text{\bf Industry} & \text{\bf Sub-Industry} \\ \hline \multicolumn{5}{c}{\text{\bf In Node Degree}} \\ \hline 113 & \text{Lincoln National Corp} & \text{USA} & \text{Insurance} & \text{Life/Health Insurance} \\ 110 & \text{Provident Financial} & \text{UK} & \text{Diversified Finan Serv} & \text{Finance-Consumer Loans} \\ 106 & \text{ING Groep NV} & \text{Netherlands} & \text{Insurance} & \text{Life/Health Insurance} \\ 105 & \text{Hartford Financial Services Group} & \text{USA} & \text{Insurance} & \text{Multi-line Insurance} \\ 103 & \text{Genworth Financial} & \text{USA} & \text{Insurance} & \text{Multi-line Insurance}\\ \hline \multicolumn{5}{c}{\text{\bf Out Node Degree}} \\ \hline 120 & \text{Provident Financial*} & \text{UK} & \text{Diversified Finan Serv*} & \text{Finance-Consumer Loans*} \\ 120 & \text{Lincoln National Corp*} & \text{USA} & \text{Insurance} & \text{Life/Health Insurance*} \\ 118 & \text{Hartford Financial Services Group*} & \text{USA} & \text{Insurance} & \text{Multi-line Insurance} \\ 113 & \text{MetLife*} & \text{USA} & \text{Insurance} & \text{Multi-line Insurance} \\ 113 & \text{Prudential Financial*} & \text{USA} & \text{Insurance} & \text{Life/Health Insurance} \\ \hline \multicolumn{5}{c}{\text{\bf In Eigenvector}} \\ \hline 9.91 & \text{ING Groep NV} & \text{Netherlands} & \text{Insurance} & \text{Life/Health Insurance} \\ 9.81 & \text{Lincoln National Corp} & \text{USA} & \text{Insurance} & \text{Life/Health Insurance} \\ 9.58 & \text{Provident Financial} & \text{UK} & \text{Diversified Finan Serv} & \text{Finance-Consumer Loans} \\ 9.57 & \text{Hartford Financial Services Group} & \text{USA} & \text{Insurance} & \text{Multi-line Insurance} \\ 9.44 & \text{Aegon NV} & \text{Netherlands} & \text{Insurance} & \text{Multi-line Insurance} \\ \hline \multicolumn{5}{c}{\text{\bf Out Eigenvector}} \\ \hline 0.126 & \text{Provident Financial*} & \text{UK} & \text{Diversified Finan Serv} & \text{Finance-Consumer Loans} \\ 0.126 & \text{Lincoln National Corp*} & \text{USA} & \text{Insurance} & \text{Life/Health Insurance} \\ 0.125 & \text{Hartford Financial Services Group*} & \text{USA} & \text{Insurance} & \text{Multi-line Insurance} \\ 0.124 & \text{MetLife*} & \text{USA} & \text{Insurance} & \text{Multi-line Insurance} \\ 0.124 & \text{Prudential Financial*} & \text{USA} & \text{Insurance} & \text{Life/Health Insurance} \\ \hline \multicolumn{5}{c}{\text{\bf In Harmonic Closeness}} \\ \hline 131.0 & \text{Provident Financial} & \text{UK} & \text{Diversified Finan Serv} & \text{Finance-Consumer Loans} \\ 129.5 & \text{Lincoln National Corp} & \text{USA} & \text{Insurance} & \text{Life/Health Insurance} \\ 127.5 & \text{Hartford Financial Services Group} & \text{USA} & \text{Insurance} & \text{Multi-line Insurance} \\ 127.0 & \text{MetLife} & \text{USA} & \text{Insurance} & \text{Multi-line Insurance} \\ 126.0 & \text{Prudential Financial} & \text{USA} & \text{Insurance} & \text{Life/Health Insurance} \\ \hline \multicolumn{5}{c}{\text{\bf Out Harmonic Closeness}} \\ \hline 134.5 & \text{Lincoln National Corp*} & \text{USA} & \text{Insurance} & \text{Life/Health Insurance} \\ 134.5 & \text{Provident Financial*} & \text{UK} & \text{Diversified Finan Serv} & \text{Finance-Consumer Loans} \\ 133.5 & \text{Genworth Financial*} & \text{USA} & \text{Insurance} & \text{Multi-line Insurance} \\ 131 & \text{Hartford Financial Services Group*} & \text{USA} & \text{Insurance} & \text{Multi-line Insurance} \\ 131 & \text{Prudential Financial*} & \text{USA} & \text{Insurance} & \text{Life/Health Insurance} \\ \hline \multicolumn{5}{c}{\text{\bf Directed Betweenness}} \\ \hline 1486 & \text{ING Groep*} & \text{Netherlands} & \text{Insurance} & \text{Life/Health Insurance} \\ 911 & \text{Lincoln National Corp*} & \text{USA} & \text{Insurance} & \text{Life/Health Insurance} \\ 802 & \text{Provident Financial*} & \text{USA} & \text{Diversified Fin. Serv.} & \text{Finance - Consumer Loans} \\ 705 & \text{Hartford Financial Services Group*} & \text{USA} & \text{Insurance} & \text{Multi-line Insurance} \\ 636 & \text{Aegon*} & \text{Netherlands} & \text{Insurance} & \text{Multi-line Insurance}\\ \hline \end{array} \]

\normalsize

\noindent {\bf Table 3.} Classification of stocks with highest centrality measures, the countries they belong to, their industry and sub-industry classifications, for asset graphs based on threshold 0.1. Only the five stocks with highest centrality values are shown (more, in case of draws).

\vskip 0.3 cm

\scriptsize

\[ \begin{array}{c|l|l|l|l} \hline \text{\bf Centrality} & \text{\bf Company} & \text{\bf Country} & \text{\bf Industry} & \text{\bf Sub-Industry} \\ \hline \multicolumn{5}{c}{\text{\bf In Node Strength}} \\ \hline 30.34 & \text{Hartford Financial Services Group} & \text{USA} & \text{Insurance} & \text{Multi-line Insurance} \\ 29.86 & \text{Lincoln National Corp} & \text{USA} & \text{Insurance} & \text{Life/Health Insurance} \\ 29.77 & \text{Prudential Financial} & \text{USA} & \text{Insurance} & \text{Life/Health Insurance} \\ 29.22 & \text{Principal Financial Group} & \text{USA} & \text{Insurance} & \text{Life/Health Insurance} \\ 27.87 & \text{Citigroup} & \text{USA} & \text{Banks} & \text{Diversified Banking Inst} \\ \hline \multicolumn{5}{c}{\text{\bf Out Node Strength}} \\ \hline 30.16 & \text{Hartford Financial Services Group *} & \text{USA} & \text{Insurance} & \text{Multi-line Insurance} \\ 28.71 & \text{Prudential Financial *} & \text{USA} & \text{Insurance} & \text{Life/Health Insurance} \\ 27.83 & \text{Lincoln National *} & \text{USA} & \text{Insurance} & \text{Life/Health Insurance} \\ 27.31 & \text{Principal Financial Group *} & \text{USA} & \text{Insurance} & \text{Life/Health Insurance} \\ 26.57 & \text{ING Groep NV *} & \text{Netherlands} & \text{Insurance} & \text{Life/Health Insurance} \\ \hline \end{array} \]

\normalsize

\noindent {\bf Table 4.} Top five stocks according to In Node Strength and to Out Node Strength, the countries they belong to, their industry and sub-industry classifications. Nodes related with lagged stocks have an asterisk beside their names. Calculations were based on the ETEs between stocks.

\vskip 0.3 cm

\section{Relations with economies in crisis}

Economic broadcasts of the past few years constantly warned of the dangers of a new global financial crisis that may be triggered by the failure of some European countries to pay their sovereign debts. It is not completely clear how far reaching a default by one of those countries could be, and which institutions are more vulnerable to that. Using networks based on financial loans and debts between banks, researchers can try to gauge some of the consequences of defaults in banks, but, as said in the introduction, networks built on loans and debts do not account for a myriad of other economical facts that define the relationships between financial institutions. So, in order to attempt to study those relations, we shall build networks based on the ETEs between the 197 major financial institutions considered until now together with all financial institutions listed in Bloomberg of some of those countries in crisis, after a liquidity filter. The aim is to investigate which of the main financial institutions receive more entropy from the financial institutions of those countries, meaning that the prices of stocks from those target institutions are much influenced by the prices of institutions that might be in danger of collapse. Of course, we are not saying here that the institutions being considered that belong to one of the countries in crisis might default; we just analyze what could happen if they did.

The countries we shall consider here are Greece, Cyprus, Ireland, Spain, Portugal, and Italy. We will do a separate analysis for each country, following the same procedures. First, we remove the stocks belonging to the country in crisis from the original network of financial institutions; then we add to this network all stocks that belong to the country in crisis and that are listed in Bloomberg. The number os stocks from each country are restrained by the data available and by the liquidity of those stocks. The second condition eliminates many of the time series available, particularly in less developed stock markets.

Greece is represented by 17 stocks, including the Bank of Greece, which is removed from the 197 original stocks of financial companies. For Cyprus, we obtain the time series of 20 stocks, after removing the less liquid ones. Spain is one of the main players in the international fears for the world economic market; we remove the stocks belonging to Spanish companies (four of them) from the bulk of main stocks and then add 26 stocks of financial companies from that country, including the ones that have been previously removed. Portugal is also an important country in the monitoring for an economic crisis since its institutions have deep connections with Spanish companies. In order to study the influence of its stocks on other stocks of main financial companies, we first remove the one stock belonging to Portugal in that group, that of the Banco Espírito Santo. Then we add to the data the log-returns of five major Portuguese banks, including the one that had been removed from the main block. The country in this group with the largest number of companies that take part of the original data set, 6 of them, is Italy, for which we start by removing those stocks from the main block, including the 6 original ones. Then we add 61 stocks belonging to the financial sector which are negotiated in Italy and which survive the liquidity filter. For Ireland, we have four stocks that survive the liquidity filter.

The Transfer Entropy, Effective Transfer Entropy, and Normalized Transfer Entropy matrices for the main block, together with the stocks belonging to each country in crisis, for each country separately, are calculated using the same techniques described in the last section. Coordinates are associated to each stock, with the reassignments slightly changing the positions of the original stocks. Figure 21 shows the stocks belonging to each country in crisis (black dots) and the stocks belonging to the other countries (white dots) in such a way that their distances represent the approximate smallest value of the Normalized Transfer Entropy between nodes $i$ and $j$ (we choose, as before, the smallest distance between $D_{ij}$ and $D_{ji}$). The nodes corresponding to the lagged stocks are not represented in the graph, and the connections between stocks are also not shown.

Stocks from Greece and Cyprus occupy positions close to the stocks of Australasia, probably a consequence of the time zones in which those two markets operate. The stocks from Spain and Portugal, Italy and Ireland are scattered along the main European cluster.

Figure 22 shows false color maps of the ETEs from lagged stocks belonging to the countries in crisis to the other stocks of major financial companies. Looking at the ETEs from Greek companies, one can see a medium transfer of entropy from those stocks mainly to stocks of European companies. The ETEs from Cypriot companies are not particularly strong, except for the stocks of some banks, which transfer entropy mainly to stocks from Europe and, particularly, to stocks from Greece. Stocks from Spain also influence mainly stocks of European financial companies. Some Portuguese stocks have large ETEs to European companies and, mainly, to some stocks from Spain. Stocks from Italy have some strong influence on stocks from other European financial companies, and stocks from Ireland have some mild influences on European stocks.

Table 5 shows the first five stocks that receive the most ETE from the stocks of each country in crisis. Almost all stocks that receive the most ETE are banks, with the exception of the ING Groep, which is a Dutch corporation that specializes in general banking services and in insurance, and so is not just an insurance company, but also a bank. The stocks that are most affected by Greek stocks are well spread among European banks, with the most affected one being the ING Grope from the Netherlands. The stock most affected by Cypriot stocks is the one of the National Bank of Greece, what is expected due to the economic and financial relations between Cyprus and Greece. The remaining influence is evenly divided by some other European stocks. The ETE transmitted from Spain to the five most influenced stocks is larger than the ETE transmitted by Greece and Cyprus, and the influence is evenly divided among the European stocks. Portuguese stocks transmit more entropy to two of the largest Spanish banks, and also to some other European stocks. The influence of Italian stocks is much larger than the influence of other stocks belonging to the group of countries in crisis, and it spreads rather evenly among some European stocks. The influence from Irish stocks is low, and evenly distributed among European stocks, including two from the UK.

One must keep in mind that what we are measuring is the sum of ETEs to a particular company, and so the number of companies that send the ETEs is important, but since the number of relevant financial companies a country has is an important factor of its influence, we here consider the sum of ETEs as a determinant of the influence of one country on another.

It is interesting to see that there are some stocks that are consistently more influenced by the stocks of countries in crisis. The Deutsche Bank appears in five lists, and the ING Groep and the KBC Groep appear in four lists. Most of the stocks listed are also some of the more central ones according to different centrality criteria.

\scriptsize

\[ \begin{array}{l|c|l|l|l} \hline \text{\bf Stock} & \text{\bf ETE} & \text{\bf Country} & \text{\bf Industry} & \text{\bf Sub-industry} \\ \hline \multicolumn{5}{c}{\text{\bf Greece}} \\ \hline \text{ING Groep} & 1.04 & \text{Netherlands} & \text{Insurance} & \text{Life/Health Insurance} \\ \text{KBC Groep} & 1.04 & \text{Belgium} & \text{Banks} & \text{Commercial Banks} \\ \text{Deutsche Bank} & 0.98 & \text{Germany} & \text{Banks} & \text{Diversified Banking Institution} \\ \text{Société Générale} & 0.98 & \text{France} & \text{Banks} & \text{Diversified Banking Institution} \\ \text{Crédit Agricole} & 0.94 & \text{France} & \text{Banks} & \text{Diversified Banking Institution} \\ \hline \multicolumn{5}{c}{\text{\bf Cyprus}} \\ \hline \text{National Bank of Greece} & 0.68& \text{Greece} & \text{Banks} & \text{Commercial Banks} \\ \text{KBC Groep NV} & 0.34& \text{Belgium} & \text{Banks} & \text{Commercial Banks} \\ \text{Deutsche Bank AG} & 0.33& \text{Germany} & \text{Banks} & \text{Diversified Banking Institution} \\ \text{ING Groep NV} & 0.30& \text{Netherlands} & \text{Insurance} & \text{Life/Health Insurance} \\ \text{DANSKE DC} & 0.28& \text{Denmark} & \text{Banks} & \text{Commercial Banks} \\ \hline \multicolumn{5}{c}{\text{\bf Spain}} \\ \hline \text{Deutsche Bank} & 2.34& \text{Germany} & \text{Banks} & \text{Diversified Banking Institution} \\ \text{BNP Paribas} & 2.33& \text{France} & \text{Banks} & \text{Diversified Banking Institution} \\ \text{AXA} & 2.31& \text{France} & \text{Insurance} & \text{Multi-line Insurance} \\ \text{ING Groep} & 2.21& \text{Netherlands} & \text{Insurance} & \text{Life/Health Insurance} \\ \text{KBC Groep} & 2.17& \text{Belgium} & \text{Banks} & \text{Commercial Bank} \\ \hline \multicolumn{5}{c}{\text{\bf Portugal}} \\ \hline \text{Banco Santander} & 0.91& \text{Spain} & \text{Banks} & \text{Commercial Bank} \\ \text{Banco Bilbao Vizcaya Argentaria} & 0.72& \text{Spain} & \text{Banks} & \text{Commercial Bank} \\ \text{BNP Paribas} & 0.62& \text{France} & \text{Banks} & \text{Diversified Banking Institution} \\ \text{Deutsche Bank} & 0.60& \text{Germany} & \text{Banks} & \text{Diversified Banking Institution} \\ \text{AXA} & 0.60& \text{France} & \text{Insurance} & \text{Multi-line Insurance} \\ \hline \multicolumn{5}{c}{\text{\bf Italy}} \\ \hline \text{AXA} & 6.37& \text{France} & \text{Insurance} & \text{Multi-line Insurance} \\ \text{Deutsche Bank AG} & 6.29& \text{Germany} & \text{Banks} & \text{Diversified Banking Institution} \\ \text{BNP Paribas} & 6.18& \text{France} & \text{Banks} & \text{Diversified Banking Institution} \\ \text{Banco Bilbao Vizcaya Argentaria} & 5.90& \text{Spain} & \text{Banks} & \text{Commercial Bank} \\ \text{Societe Generale} & 5.84& \text{France} & \text{Banks} & \text{Diversified Banking Institution} \\ \hline \multicolumn{5}{c}{\text{\bf Ireland}} \\ \hline \text{ING Groep NV} & 0.39& \text{Netherlands} & \text{Insurance} & \text{Life/Health Insurance} \\ \text{Barclays} & 0.37& \text{UK} & \text{Banks} & \text{Diversified Banking Institution} \\ \text{Lloyds Banking Group} & 0.37& \text{UK} & \text{Banks} & \text{Diversified Banking Institution} \\ \text{Aegon NV} & 0.36& \text{Netherlands} & \text{Insurance} & \text{Multi-line Insurance} \\ \text{KBC Groep NV} & 0.36& \text{Belgium} & \text{Banks} & \text{Commercial Bank} \\ \hline \end{array} \]

\normalsize

\noindent {\bf Table 5.} Five stocks that receive more ETE from the stocks of each country in crisis. In the table, are shown the name of the company, the total ETE received from the stocks of countries in crisis, the country the stock belongs to, the industry and sub-industry.

\vskip 0.3 cm

Table 6 shows the first five stocks that send the most ETE from the stocks of each country in crisis (four, in the case of Ireland). The most influential stocks are mainly those of banks, but we also have highly influent stocks belonging to insurance companies and to investment companies. The influence of Greece is distributed among some banks, and the influence of Cyprus is also mainly distributed among banks. The Spanish influence also comes from commercial banks, and is concentrated on the top three ones. The same applies to Portugal, with the main ETE being transmitted from a stock that belongs to a Spanish bank but that is also negotiated in Portugal. The most influential stocks from Italy are those of companies that are originally from other European countries, but whose stocks are also negotiated in Italy. The influence of Ireland is mainly distributed among two banks and one insurance company.

So we may conclude that the most influenced stocks by stocks of the countries in crisis according to ETE are those of European companies, and mainly some stocks belonging to some particular banks. The stocks that influence the most, also according to the ETE criterium, are those of banks belonging to the countries in crisis, in particular if the banks are native to other countries, but their stocks are negotiated in the country in crisis.

\scriptsize

\[ \begin{array}{l|c|l|l} \hline \text{\bf Stock} & \text{\bf ETE} & \text{\bf Industry} & \text{\bf Sub-industry} \\ \hline \multicolumn{4}{c}{\text{\bf Greece}} \\ \hline \text{National Bank of Greece} & 5.95 & \text{Banks} & \text{Commercial Bank} \\ \text{Piraeus Bank} & 4.68 & \text{Banks} & \text{Commercial Bank} \\ \text{Cyprus Popular Bank} & 4.48 & \text{Banks} & \text{Commercial Bank} \\ \text{Eurobank Ergasias} & 4.38 & \text{Banks} & \text{Commercial Bank} \\ \text{Bank of Cyprus} & 4.28 & \text{Banks} & \text{Commercial Bank} \\ \hline \multicolumn{4}{c}{\text{\bf Cyprus}} \\ \hline \text{Cyprus Popular Bank} & 5.18 & \text{Banks} & \text{Commercial Banks} \\ \text{Bank of Cyprus} & 4.01 & \text{Banks} & \text{Commercial Banks} \\ \text{Hellenic Bank} & 3.02 & \text{Banks} & \text{Commercial Banks} \\ \text{Interfund Investments} & 2.12 & \text{Investment Companies} & \text{Investment Companies} \\ \text{Demetra Investments} & 1.88 & \text{Investment Companies} & \text{Investment Companies} \\ \hline \multicolumn{4}{c}{\text{\bf Spain}} \\ \hline \text{Banco Santander} & 15.90 & \text{Banks} & \text{Commercial Bank} \\ \text{Banco Bilbao Vizcaya Argentaria} & 14.74 & \text{Bank} & \text{Commercial Bank} \\ \text{Banco Popular Espanol} & 11.35 & \text{Banks} & \text{Commercial Bank} \\ \text{Banco de Sabadell} & 10.47 & \text{Bank} & \text{Commercial Bank} \\ \text{Banco Bradesco} & 9.99 & \text{Banks} & \text{Commercial Bank} \\ \hline \multicolumn{4}{c}{\text{\bf Portugal}} \\ \hline \text{Banco Santander} & 12.67 & \text{Banks} & \text{Commercial Banks} \\ \text{Banco Espírito Santo} & 8.60 & \text{Banks} & \text{Commercial Banks} \\ \text{Banco BPI} & 8.32 & \text{Banks} & \text{Commercial Banks} \\ \text{Banco Comercial Portugues} & 3.08 & \text{Banks} & \text{Commercial Banks} \\ \text{Espírito Santo Financial Group} & 4.08 & \text{Banks} & \text{Commercial Banks} \\ \hline \multicolumn{4}{c}{\text{\bf Italy}} \\ \hline \text{ING Groep NV} & 15.91 & \text{Insurance} & \text{Life - Health Insurance} \\ \text{Deutsche Bank AG} & 15.43 & \text{Banks} & \text{Diversified Banking Institution} \\ \text{AXA} & 15.23 & \text{Insurance} & \text{Multi-line Insurance} \\ \text{BNP Paribas} & 14.51 & \text{Banks} & \text{Diversified Banking Institution} \\ \text{UniCredit SpA} & 14.09 & \text{Banks} & \text{Diversified Banking Institution} \\ \hline \multicolumn{4}{c}{\text{\bf Ireland}} \\ \hline \text{Bank of Ireland} & 12.67 & \text{Banks} & \text{Commercial Bank} \\ \text{Permanent TSB Group Holdings} & 8.60 & \text{Insurance} & \text{Property - Casualty Insurance} \\ \text{Allied Irish Banks} & 8.32 & \text{Banks} & \text{Commercial Bank} \\ \text{FBD Holdings} & 3.08 & \text{Insurance} & \text{Property - Casualty Insurance} \\ \hline \end{array} \]

\normalsize

\noindent {\bf Table 6.} Five stocks that send more ETE from each country in crisis. In the table, are shown the name of the company, the total ETE sent to the stocks of main financial companies, the industry and sub-industry.

\vskip 0.3 cm

\section{Conclusions}

We have seen in this work how the stocks of the top 197 financial companies, in market volume, relate to one another, using both their correlations and the Transfer Entropy between them. We saw that they are related first by country where the stocks are negotiated, and then by industry and sub-industry. The network structures for correlation and for Transfer Entropy are very different from one another, being the network obtained using Transfer Entropy a directed one, with causal influences between the stocks. The use of original and lagged log-returns also revealed some relationships between stocks, with the stocks of a previous day influencing the stocks of the following day. A study of the centralities of the stocks revealed that the most central ones are those of insurance companies of Europe and of the USA, or of banks of the USA and Europe. Since insurance and reinsurance companies are major CDS (Credit Default Securities) sellers, and banks are both major CDS buyers and sellers, some of this centrality of insurance companies, followed by banks, might be explained by the selling and buying of CDS.

A further study of the causality relations between stocks of companies belonging to countries in crisis, namely Greece, Cyprus, Spain, Portugal, Italy, and Ireland, reveal which are the most affected financial companies belonging to the group of largest financial stocks. This calls attention to liabilities of those companies to possible defaults or fall of stocks prices of companies belonging to those countries in crisis.

This work plants the seeds for the study of contagion among financial institutions, but now based on a real network, showing which companies are most central for the propagation of crises and which ones are more dependent on failing economies. This may be used to develop policies for avoiding the spread of financial crises.

\vskip 0.6 cm

\noindent{\bf Acknowledgements}

\vskip 0.4 cm

The author acknowledges the support of this work by a grant from Insper, Instituto de Ensino e Pesquisa. This article was written using \LaTeX, all figures were made using PSTricks and Matlab, and the calculations were made using Matlab and Excel. Special thanks to Raul Ikeda, who determined and collected the data, and to Nanci Romero, who helped organizing and classifying the data according to industry and sub-industry.

\appendix

\section{List of stocks used}

Here are displayed, in order of country and of industry and sub-industry, the stocks that are used in the present work, not considering stocks from particular countries in crisis.

\vskip 0.3 cm

\scriptsize

\[ \begin{array}{l|l|l|l} \hline \text{Country} & \text{Company} & \text{Industry} & \text{Sector} \\ \hline
\text{USA 1} & \text{Bank of America Corp} & \text{Banks} & \text{Diversified Banking Inst} \\
\text{USA 2} & \text{Citigroup Inc} & \text{Banks} & \text{Diversified Banking Inst} \\
\text{USA 3} & \text{Goldman Sachs Group Inc/The} & \text{Banks} & \text{Diversified Banking Inst} \\
\text{USA 4} & \text{JPMorgan Chase \& Co} & \text{Banks} & \text{Diversified Banking Inst} \\
\text{USA 5} & \text{Morgan Stanley} & \text{Banks} & \text{Diversified Banking Inst} \\
\text{USA 6} & \text{Comerica Inc} & \text{Banks} & \text{Super-Regional Banks-US} \\
\text{USA 7} & \text{Capital One Financial Corp} & \text{Banks} & \text{Super-Regional Banks-US} \\
\text{USA 8} & \text{KeyCorp} & \text{Banks} & \text{Super-Regional Banks-US} \\
\text{USA 9} & \text{PNC Financial Services Group Inc/The} & \text{Banks} & \text{Super-Regional Banks-US} \\
\text{USA 10} & \text{SunTrust Banks Inc} & \text{Banks} & \text{Super-Regional Banks-US} \\
\text{USA 11} & \text{US Bancorp} & \text{Banks} & \text{Super-Regional Banks-US} \\
\text{USA 12} & \text{Wells Fargo \& Co} & \text{Banks} & \text{Super-Regional Banks-US} \\
\text{USA 13} & \text{Fifth Third Bancorp} & \text{Banks} & \text{Super-Regional Banks-US} \\
\text{USA 14} & \text{Huntington Bancshares Inc/OH} & \text{Banks} & \text{Super-Regional Banks-US} \\
\text{USA 15} & \text{BB\&T Corp} & \text{Banks} & \text{Commer Banks-Southern US} \\
\text{USA 16} & \text{First Horizon National Corp} & \text{Banks} & \text{Commer Banks-Southern US} \\
\text{USA 17} & \text{Regions Financial Corp} & \text{Banks} & \text{Commer Banks-Southern US} \\
\text{USA 18} & \text{M\&T Bank Corp} & \text{Banks} & \text{Commer Banks-Eastern US} \\
\text{USA 19} & \text{Zions Bancorporation} & \text{Banks} & \text{Commer Banks-Western US} \\
\text{USA 20} & \text{Bank of New York Mellon Corp/The} & \text{Banks} & \text{Fiduciary Banks} \\
\text{USA 21} & \text{State Street Corp} & \text{Banks} & \text{Fiduciary Banks} \\
\text{USA 22} & \text{Northern Trust Corp} & \text{Banks} & \text{Fiduciary Banks} \\
\text{USA 23} & \text{Banco Bradesco SA} & \text{Banks} & \text{Commer Banks Non-US} \\
\text{USA 24} & \text{Itau Unibanco Holding SA} & \text{Banks} & \text{Commer Banks Non-US} \\
\text{USA 25} & \text{Banco Santander Chile} & \text{Banks} & \text{Commer Banks Non-US} \\
\text{USA 26} & \text{Credicorp Ltd} & \text{Banks} & \text{Commer Banks Non-US} \\
\text{USA 27} & \text{American Express Co} & \text{Diversified Finan Serv} & \text{Finance-Credit Card} \\
\text{USA 28} & \text{Ameriprise Financial Inc} & \text{Diversified Finan Serv} & \text{Invest Mgmnt/Advis Serv} \\
\text{USA 29} & \text{Franklin Resources Inc} & \text{Diversified Finan Serv} & \text{Invest Mgmnt/Advis Serv} \\
\text{USA 30} & \text{BlackRock Inc} & \text{Diversified Finan Serv} & \text{Invest Mgmnt/Advis Serv} \\
\text{USA 31} & \text{Invesco Ltd} & \text{Diversified Finan Serv} & \text{Invest Mgmnt/Advis Serv} \\
\text{USA 32} & \text{Legg Mason Inc} & \text{Diversified Finan Serv} & \text{Invest Mgmnt/Advis Serv} \\
\text{USA 33} & \text{T Rowe Price Group Inc} & \text{Diversified Finan Serv} & \text{Invest Mgmnt/Advis Serv} \\
\text{USA 34} & \text{E*TRADE Financial Corp} & \text{Diversified Finan Serv} & \text{Finance-Invest Bnkr/Brkr} \\
\text{USA 35} & \text{IntercontinentalExchange Inc} & \text{Diversified Finan Serv} & \text{Finance-Other Services} \\
\text{USA 36} & \text{NYSE Euronext} & \text{Diversified Finan Serv} & \text{Finance-Other Services} \\
\hline \end{array} \]

\[ \begin{array}{l|l|l|l} \hline \text{Country} & \text{Company} & \text{Industry} & \text{Sector} \\ \hline
\text{USA 37} & \text{NASDAQ OMX Group Inc/The} & \text{Diversified Finan Serv} & \text{Finance-Other Services} \\
\text{USA 38} & \text{Hudson City Bancorp Inc} & \text{Savings \& Loans} & \text{S\& L/Thrifts-Eastern US} \\
\text{USA 39} & \text{People's United Financial Inc} & \text{Savings \& Loans} & \text{S\& L/Thrifts-Eastern US} \\
\text{USA 40} & \text{ACE Ltd} & \text{Insurance} & \text{Multi-line Insurance} \\
\text{USA 41} & \text{American International Group Inc} & \text{Insurance} & \text{Multi-line Insurance} \\
\text{USA 42} & \text{Assurant Inc} & \text{Insurance} & \text{Multi-line Insurance} \\
\text{USA 43} & \text{Allstate Corp/The} & \text{Insurance} & \text{Multi-line Insurance} \\
\text{USA 44} & \text{Genworth Financial Inc} & \text{Insurance} & \text{Multi-line Insurance} \\
\text{USA 45} & \text{Hartford Financial Services Group Inc} & \text{Insurance} & \text{Multi-line Insurance} \\
\text{USA 46} & \text{Loews Corp} & \text{Insurance} & \text{Multi-line Insurance} \\
\text{USA 47} & \text{MetLife Inc} & \text{Insurance} & \text{Multi-line Insurance} \\
\text{USA 48} & \text{XL Group PLC} & \text{Insurance} & \text{Multi-line Insurance} \\
\text{USA 49} & \text{Cincinnati Financial Corp} & \text{Insurance} & \text{Multi-line Insurance} \\
\text{USA 50} & \text{Principal Financial Group Inc} & \text{Insurance} & \text{Life/Health Insurance} \\
\text{USA 51} & \text{Lincoln National Corp} & \text{Insurance} & \text{Life/Health Insurance} \\
\text{USA 52} & \text{Aflac Inc} & \text{Insurance} & \text{Life/Health Insurance} \\
\text{USA 53} & \text{Torchmark Corp} & \text{Insurance} & \text{Life/Health Insurance} \\
\text{USA 54} & \text{Unum Group} & \text{Insurance} & \text{Life/Health Insurance} \\
\text{USA 55} & \text{Prudential Financial Inc} & \text{Insurance} & \text{Life/Health Insurance} \\
\text{USA 56} & \text{Travelers Cos Inc/The} & \text{Insurance} & \text{Property/Casualty Ins} \\
\text{USA 57} & \text{Chubb Corp/The} & \text{Insurance} & \text{Property/Casualty Ins} \\
\text{USA 58} & \text{Progressive Corp/The} & \text{Insurance} & \text{Property/Casualty Ins} \\
\text{USA 59} & \text{Aon PLC} & \text{Insurance} & \text{Insurance Brokers} \\
\text{USA 60} & \text{Marsh \& McLennan Cos Inc} & \text{Insurance} & \text{Insurance Brokers} \\
\text{USA 61} & \text{Berkshire Hathaway Inc} & \text{Insurance} & \text{Reinsurance} \\
\text{USA 62} & \text{CBRE Group Inc} & \text{Real Estate} & \text{Real Estate Mgmnt/Servic} \\
\text{USA 63} & \text{Apartment Investment \& Management Co} & \text{REITS} & \text{REITS-Apartments} \\
\text{USA 64} & \text{AvalonBay Communities Inc} & \text{REITS} & \text{REITS-Apartments} \\
\text{USA 65} & \text{Equity Residential} & \text{REITS} & \text{REITS-Apartments} \\
\text{USA 66} & \text{Boston Properties Inc} & \text{REITS} & \text{REITS-Office Property} \\
\text{USA 67} & \text{Host Hotels \& Resorts Inc} & \text{REITS} & \text{REITS-Hotels} \\
\text{USA 68} & \text{Prologis Inc} & \text{REITS} & \text{REITS-Warehouse/Industr} \\
\text{USA 69} & \text{Public Storage} & \text{REITS} & \text{REITS-Storage} \\
\text{USA 70} & \text{Simon Property Group Inc} & \text{REITS} & \text{REITS-Regional Malls} \\
\text{USA 71} & \text{Macerich Co/The} & \text{REITS} & \text{REITS-Regional Malls} \\
\text{USA 72} & \text{Kimco Realty Corp} & \text{REITS} & \text{REITS-Shopping Centers} \\
\text{USA 73} & \text{Ventas Inc} & \text{REITS} & \text{REITS-Health Care} \\
\text{USA 74} & \text{HCP Inc} & \text{REITS} & \text{REITS-Health Care} \\
\text{USA 75} & \text{Health Care REIT Inc} & \text{REITS} & \text{REITS-Health Care} \\
\text{USA 76} & \text{American Tower Corp} & \text{REITS} & \text{REITS-Diversified} \\
\text{USA 77} & \text{Weyerhaeuser Co} & \text{REITS} & \text{REITS-Diversified} \\
\text{USA 78} & \text{Vornado Realty Trust} & \text{REITS} & \text{REITS-Diversified} \\
\text{USA 79} & \text{Plum Creek Timber Co Inc} & \text{REITS} & \text{REITS-Diversified} \\
\text{Canada 1} & \text{Bank of Montreal} & \text{Banks} & \text{Commer Banks Non-US} \\
\text{Canada 2} & \text{Bank of Nova Scotia} & \text{Banks} & \text{Commer Banks Non-US} \\
\text{Canada 3} & \text{Canadian Imperial Bank of Commerce/Canada} & \text{Banks} & \text{Commer Banks Non-US} \\
\text{Canada 4} & \text{National Bank of Canada} & \text{Banks} & \text{Commer Banks Non-US} \\
\text{Canada 5} & \text{Royal Bank of Canada} & \text{Banks} & \text{Commer Banks Non-US} \\
\text{Canada 6} & \text{Toronto-Dominion Bank/The} & \text{Banks} & \text{Commer Banks Non-US} \\
\text{Canada 7} & \text{Manulife Financial Corp} & \text{Insurance} & \text{Life/Health Insurance} \\
\text{Canada 8} & \text{Power Corp of Canada} & \text{Insurance} & \text{Life/Health Insurance} \\
\text{Canada 9} & \text{Sun Life Financial Inc} & \text{Insurance} & \text{Life/Health Insurance} \\
\text{Canada 10} & \text{Brookfield Asset Management Inc} & \text{Real Estate} & \text{ Real Estate Oper/Develop} \\
\text{Chile} & \text{Banco de Chil} & \text{Banks} & \text{Commer Banks Non-US} \\
\text{UK 1} & \text{Barclays PLC} & \text{Banks} & \text{Diversified Banking Inst} \\
\text{UK 2} & \text{HSBC Holdings PLC} & \text{Banks} & \text{Diversified Banking Inst} \\
\text{UK 3} & \text{Lloyds Banking Group PLC} & \text{Banks} & \text{Diversified Banking Inst} \\
\text{UK 4} & \text{Royal Bank of Scotland Group PLC} & \text{Banks} & \text{Diversified Banking Inst} \\
\text{UK 5} & \text{Standard Chartered PLC} & \text{Banks} & \text{Commer Banks Non-US} \\
\text{UK 6} & \text{Aberdeen Asset Management PLC} & \text{Diversified Finan Serv} & \text{Invest Mgmnt/Advis Serv} \\
\text{UK 7} & \text{Man Group PLC} & \text{Diversified Finan Serv} & \text{Invest Mgmnt/Advis Serv} \\
\text{UK 8} & \text{Schroders PLC} & \text{Diversified Finan Serv} & \text{Invest Mgmnt/Advis Serv} \\
\text{UK 9} & \text{Old Mutual PLC} & \text{Diversified Finan Serv} & \text{Invest Mgmnt/Advis Serv} \\
\text{UK 10} & \text{Provident Financial PLC} & \text{Diversified Finan Serv} & \text{Finance-Consumer Loans} \\
\text{UK 11} & \text{London Stock Exchange Group PLC} & \text{Diversified Finan Serv} & \text{Finance-Other Services} \\
\text{UK 12} & \text{Aviva PLC} & \text{Insurance} & \text{Life/Health Insurance} \\
\text{UK 13} & \text{Legal \& General Group PLC} & \text{Insurance} & \text{Life/Health Insurance} \\
\text{UK 14} & \text{Prudential PLC} & \text{Insurance} & \text{Life/Health Insurance} \\
\text{UK 15} & \text{Standard Life PLC} & \text{Insurance} & \text{Life/Health Insurance} \\
\text{UK 16} & \text{RSA Insurance Group PLC} & \text{Insurance} & \text{Property/Casualty Ins} \\
\text{UK 17} & \text{3i Group PLC} & \text{Private} & \text{Private} \\
\text{UK 18} & \text{Hammerson PLC} & \text{REITS} & \text{REITS-Shopping Centers} \\
\hline \end{array} \]

\[ \begin{array}{l|l|l|l} \hline \text{Country} & \text{Company} & \text{Industry} & \text{Sector} \\ \hline
\text{UK 19} & \text{British Land Co PLC} & \text{REITS} & \text{REITS-Diversified} \\
\text{UK 20} & \text{Land Securities Group PLC} & \text{REITS} & \text{REITS-Diversified} \\
\text{UK 21} & \text{Segro PLC} & \text{REITS} & \text{REITS-Diversified} \\
\text{France 1} & \text{Credit Agricole SA} & \text{Banks} & \text{Diversified Banking Inst} \\
\text{France 2} & \text{BNP Paribas SA} & \text{Banks} & \text{Diversified Banking Inst} \\
\text{France 3} & \text{Societe Generale SA} & \text{Banks} & \text{Diversified Banking Inst} \\
\text{France 4} & \text{AXA SA} & \text{Insurance} & \text{Multi-line Insurance} \\
\text{Germany 1} & \text{Commerzbank AG} & \text{Banks} & \text{Commer Banks Non-US} \\
\text{Germany 2} & \text{Deutsche Bank AG} & \text{Banks} & \text{Diversified Banking Inst} \\
\text{Germany 3} & \text{Deutsche Boerse AG} & \text{Diversified Finan Serv} & \text{Finance-Other Services} \\
\text{Germany 4} & \text{Allianz SE} & \text{Insurance} & \text{Multi-line Insurance} \\
\text{Germany 5} & \text{Muenchener Rueckversicherungs AG} & \text{Insurance} & \text{Reinsurance} \\
\text{Switzerland 1} & \text{Credit Suisse Group AG} & \text{Banks} & \text{Diversified Banking Inst} \\
\text{Switzerland 2} & \text{UBS AG} & \text{Banks} & \text{Diversified Banking Inst} \\
\text{Switzerland 3} & \text{GAM Holding AG} & \text{Diversified Finan Serv} & \text{Invest Mgmnt/Advis Serv} \\
\text{Switzerland 4} & \text{Baloise Holding AG} & \text{Insurance} & \text{Multi-line Insurance} \\
\text{Switzerland 5} & \text{Zurich Insurance Group AG} & \text{Insurance} & \text{Multi-line Insurance} \\
\text{Switzerland 6} & \text{Swiss Life Holding AG} & \text{Insurance} & \text{Life/Health Insurance} \\
\text{Switzerland 7} & \text{Swiss Re AG} & \text{Insurance} & \text{Reinsurance} \\
\text{Austria} & \text{Erste Group Bank AG} & \text{Banks} & \text{Commer Banks Non-US} \\
\text{Netherlands 1} & \text{Aegon NV} & \text{Insurance} & \text{Multi-line Insurance} \\
\text{Netherlands 2} & \text{ING Groep NV} & \text{Insurance} & \text{Life/Health Insurance} \\
\text{Belgium 1} & \text{KBC Groep NV} & \text{Banks} & \text{Commer Banks Non-US} \\
\text{Belgium 2} & \text{Ageas} & \text{Insurance} & \text{Life/Health Insurance} \\
\text{Sweden 1} & \text{Nordea Bank AB} & \text{Banks} & \text{Commer Banks Non-US} \\
\text{Sweden 2} & \text{Skandinaviska Enskilda Banken AB} & \text{Banks} & \text{Commer Banks Non-US} \\
\text{Sweden 3} & \text{Svenska Handelsbanken AB} & \text{Banks} & \text{Commer Banks Non-US} \\
\text{Sweden 4} & \text{Swedbank AB} & \text{Banks} & \text{Commer Banks Non-US} \\
\text{Sweden 5} & \text{Investor AB} & \text{Investment Companies} & \text{Investment Companies} \\
\text{Denmark} & \text{Danske Bank A/S} & \text{Banks} & \text{Commer Banks Non-US} \\
\text{Finland} & \text{Sampo} & \text{Insurance} & \text{Multi-line Insurance} \\
\text{Norway} & \text{DNB ASA} & \text{Banks} & \text{Commer Banks Non-US} \\
\text{Italy 1} & \text{Banca Monte dei Paschi di Siena SpA} & \text{Banks} & \text{Commer Banks Non-US} \\
\text{Italy 2} & \text{Intesa Sanpaolo SpA} & \text{Banks} & \text{Commer Banks Non-US} \\
\text{Italy 3} & \text{Mediobanca SpA} & \text{Banks} & \text{Commer Banks Non-US} \\
\text{Italy 4} & \text{Unione di Banche Italiane SCPA} & \text{Banks} & \text{Commer Banks Non-US} \\
\text{Italy 5} & \text{UniCredit SpA} & \text{Banks} & \text{Diversified Banking Inst} \\
\text{Italy 6} & \text{Assicurazioni Generali SpA} & \text{Insurance} & \text{Multi-line Insurance} \\
\text{Spain 1} & \text{Banco Bilbao Vizcaya Argentaria SA} & \text{Banks} & \text{Commer Banks Non-US} \\
\text{Spain 2} & \text{Banco Popular Espanol SA} & \text{Banks} & \text{Commer Banks Non-US} \\
\text{Spain 3} & \text{Banco de Sabadell SA} & \text{Banks} & \text{Commer Banks Non-US} \\
\text{Spain 4} & \text{Banco Santander SA} & \text{Banks} & \text{Commer Banks Non-US} \\
\text{Portugal} & \text{Banco Espírito Santo SA} & \text{Banks} & \text{Commer Banks Non-US} \\
\text{Greece} & \text{National Bank of Greece SA} & \text{Banks} & \text{Commer Banks Non-US} \\
\text{Japan 1} & \text{Shinsei Bank Ltd} & \text{Banks} & \text{Commer Banks Non-US} \\
\text{Japan 2} & \text{Mitsubishi UFJ Financial Group Inc} & \text{Banks} & \text{Diversified Banking Inst} \\
\text{Japan 3} & \text{Sumitomo Mitsui Trust Holdings Inc} & \text{Banks} & \text{Commer Banks Non-US} \\
\text{Japan 4} & \text{Sumitomo Mitsui Financial Group Inc} & \text{Banks} & \text{Commer Banks Non-US} \\
\text{Japan 5} & \text{Mizuho Financial Group Inc} & \text{Banks} & \text{Commer Banks Non-US} \\
\text{Japan 6} & \text{Credit Saison Co Ltd} & \text{Diversified Finan Serv} & \text{Finance-Credit Card} \\
\text{Japan 7} & \text{Daiwa Securities Group Inc} & \text{Diversified Finan Serv} & \text{Finance-Invest Bnkr/Brkr} \\
\text{Japan 8} & \text{Nomura Holdings Inc} & \text{Diversified Finan Serv} & \text{Finance-Invest Bnkr/Brkr} \\
\text{Japan 9} & \text{ORIX Corp} & \text{Diversified Finan Serv} & \text{Finance-Leasing Compan} \\
\text{Japan 10} & \text{Tokio Marine Holdings In} & \text{Insurance} & \text{Property/Casualty Ins} \\
\text{Japan 11} & \text{Mitsui Fudosan Co Ltd} & \text{Real Estate} & \text{Real Estate Oper/Develop} \\
\text{Japan 12} & \text{Mitsubishi Estate Co Ltd} & \text{Real Estate} & \text{Real Estate Mgmnt/Servic} \\
\text{Japan 13} & \text{Sumitomo Realty \& Development Co Ltd} & \text{Real Estate} & \text{Real Estate Oper/Develop} \\
\text{Hong Kong 1} & \text{Hang Seng Bank Ltd} & \text{Banks} & \text{Commer Banks Non-US} \\
\text{Hong Kong 2} & \text{Industrial \& Commercial Bank of China Ltd} & \text{Banks} & \text{Commer Banks Non-US} \\
\text{Hong Kong 3} & \text{BOC Hong Kong Holdings Ltd} & \text{Banks} & \text{Commer Banks Non-US} \\
\text{Hong Kong 4} & \text{China Construction Bank Corp} & \text{Banks} & \text{Commer Banks Non-US} \\
\text{Hong Kong 5} & \text{Hong Kong Exchanges and Clearing Ltd} & \text{Diversified Finan Serv} & \text{Finance-Other Services} \\
\text{Hong Kong 6} & \text{Ping An Insurance Group Co of China Ltd} & \text{Insurance} & \text{Multi-line Insurance} \\
\text{Hong Kong 7} & \text{China Life Insurance Co Ltd} & \text{Insurance} & \text{Life/Health Insurance} \\
\text{Hong Kong 8} & \text{Cheung Kong Holdings Ltd} & \text{Real Estate} & \text{Real Estate Oper/Develop} \\
\text{Hong Kong 9} & \text{Sun Hung Kai Properties Ltd} & \text{Real Estate} & \text{Real Estate Oper/Develop} \\
\text{South Korea} & \text{Shinhan Financial Group Co Ltd} & \text{Diversified Finan Serv} & \text{Diversified Finan Serv} \\
\text{Taiwan} & \text{Cathay Financial Holding Co Ltd} & \text{Insurance} & \text{Life/Health Insurance} \\
\text{Singapore 1} & \text{DBS Group Holdings Ltd} & \text{Banks} & \text{Commer Banks Non-US} \\
\text{Singapore 2} & \text{Oversea-Chinese Banking Corp Ltd} & \text{Banks} & \text{Commer Banks Non-US} \\
\text{Singapore 3} & \text{United Overseas Bank Ltd} & \text{Banks} & \text{Commer Banks Non-US} \\
\hline \end{array} \]

\[ \begin{array}{l|l|l|l} \hline \text{Country} & \text{Company} & \text{Industry} & \text{Sector} \\ \hline
\text{Australia 1} & \text{Australia \& New Zealand Banking Group Ltd} & \text{Banks} & \text{Commer Banks Non-US} \\
\text{Australia 2} & \text{Commonwealth Bank of Australia} & \text{Banks} & \text{Commer Banks Non-US} \\
\text{Australia 3} & \text{National Australia Bank Ltd} & \text{Banks} & \text{Commer Banks Non-US} \\
\text{Australia 4} & \text{Westpac Banking Corp} & \text{Banks} & \text{Commer Banks Non-US} \\
\text{Australia 5} & \text{Macquarie Group Ltd} & \text{Diversified Finan Serv} & \text{Finance-Invest Bnkr/Brkr} \\
\text{Australia 6} & \text{ASX Ltd} & \text{Diversified Finan Serv} & \text{Finance-Other Services} \\
\text{Australia 7} & \text{AMP Ltd} & \text{Insurance} & \text{Life/Health Insurance} \\
\text{Australia 8} & \text{Suncorp Group Ltd} & \text{Insurance} & \text{Life/Health Insurance} \\
\text{Australia 9} & \text{Insurance Australia Group Ltd} & \text{Insurance} & \text{Property/Casualty Ins} \\
\text{Australia 10} & \text{QBE Insurance Group Ltd} & \text{Insurance} & \text{Property/Casualty Ins} \\
\text{Australia 11} & \text{Lend Lease Group} & \text{Real Estate} & \text{Real Estate Mgmnt/Servic} \\
\text{Australia 12} & \text{CFS Retail Property Trust Group} & \text{REITS} & \text{REITS-Shopping Centers} \\
\text{Australia 13} & \text{Westfield Group} & \text{REITS} & \text{REITS-Shopping Centers} \\
\text{Australia 14} & \text{Dexus Property Group} & \text{REITS} & \text{REITS-Diversified} \\
\text{Australia 15} & \text{Goodman Group} & \text{REITS} & \text{REITS-Diversified} \\
\text{Australia 16} & \text{GPT Group} & \text{REITS} & \text{REITS-Diversified} \\
\text{Australia 17} & \text{Mirvac Group} & \text{REITS} & \text{REITS-Diversified} \\
\text{Australia 18} & \text{Stockland} & \text{REITS} & \text{REITS-Diversified} \\
\hline \end{array} \]

\normalsize

\vskip 0.5 cm

\noindent {\bf \Large References}

\vskip 0.5 cm

\noindent  Allen, F. and Gale, D. (2000). Financial contagion. Journal of Political Economy 108, 1-33.

\vskip 0.2 cm

\noindent Allen, F. and Babus, A. (2009). Networks in finance. In Kleindorfer, P., Wing, Y.,and Gunther, R. (eds.), The network challenge: strategy, profit, and risk in an interlinked world. Wharton School Publishing.

\vskip 0.2 cm

\noindent Ausloos, M. and Lambiotte, R. (2007). Clusters or networks of economies? A macroeconomy study through gross domestic product. Physica A 382, 16-21.

\vskip 0.2 cm

\noindent Borg, I. and Groenen, P. (2005). Modern Multidimensional Scaling: theory and applications. 2nd edition, Springer-Verlag.

\vskip 0.2 cm

\noindent Dimp, T. Huergo, L. and Peter, F.J. (2012). Using transfer entropy to measure information flows from and to the CDS market. Midwest Finance Association 2012 Annual Meetings Paper.

\vskip 0.2 cm

\noindent Haldane, A. (2009). Rethinking the financial network. Speech delivered at the Financial Student Association, Amsterdam, Aprol.

\vskip 0.2 cm

\noindent Jizba, P., Kleinert, H. and Shefaat, M. (2012). Renyi's information transfer between financial time series. Physica A 391, 2971-2989.

\vskip 0.2 cm

\noindent Kwon, O. and Yang, J-S (2008a). Information flow between composite stock index and individual stocks. Physica A 387, 2851-2856.

\vskip 0.2 cm

\noindent Kwon, O. and Yang, J-S (2008b). Information flow between stock indices. European Physics Letters 82, 68003.

\vskip 0.2 cm

\noindent Mantegna, R.N. (1999). Hierarchical structure in financial markets. The European Physics Journal B 11, 193.

\vskip 0.2 cm

\noindent Newman, M.E.J. (2010). Networks, and introduction. Oxford University Press.

\vskip 0.2 cm

\noindent Onnela, J.-P., Chakraborti, A., Kaski, K., and Kertész, J. (2002). Dynamic asset trees and portfolio analysis. The European Physics Journal B 30, 285-288.

\vskip 0.2 cm

\noindent Onnela, J.-P., Chakraborti, A., and Kaski, K. (2003). Dynamics of market correlations: taxonomy and portfolio analysis. Physical Review E 68, 1-12.

\vskip 0.2 cm

\noindent Onnela, J.-P., Chakraborti, A., Kaski, K., and Kertész, J. (2003). Dynamic asset trees and Black Monday. Physica A 324, 247-252.

\vskip 0.2 cm

\noindent Onnela, J.-P., Chakraborti, A., Kaski, K., Kertész, J., and Kanto, A. (2003). Asset trees and asset graphs in financial markets. Physica Scripta T 106, 48-54.

\vskip 0.2 cm

\noindent Onnela, J.-P., Chakraborti, A., Kaski, K., and Kertész, J. (2004). Clustering and information in correlation based financial networks. The European Physics Journal B 38, 353-362.

\vskip 0.2 cm

\noindent Sandoval Jr., L. (2012a). To lag or not to lag? How to compare indices of stock markets that operate at different times. arXiv:1201.4586.

\vskip 0.2 cm

\noindent Sandoval Jr., L. (2012b). A Map of the Brazilian Stock Market. Advances in Complex Systems 15, 1250042-1250082.

\vskip 0.2 cm

\noindent Sandoval Jr., L. (2013). Cluster formation and evolution in networks of financial market indices. Algorithmic Finance 2, 3-43.

\vskip 0.2 cm

\noindent Schreiber, T. (2000). Measuring information transfer. Physical Review Letters 85, 461-464.

\vskip 0.2 cm

\noindent Shannon, C.E. (1948). A Mathematical Theory of Communication. Bell System Technical Journal 27, 379–423, 623-656.

\vskip 0.2 cm

\noindent Sinha, S. and Pan R.K. (2007). Uncovering the internal structure of the Indian financial market: cross-correlation behavior in the NSE. In ``Econophysics of markets and business networks'', Springer, 215-226.

\vskip 0.2 cm

\noindent Upper, C. (2011). Simulation methods to assess the danger of contagion in interbank markets. Journal of Financial Stability 7 (3), 111–125.

\begin{figure}[H]
\begin{center}
\includegraphics[scale=0.85]{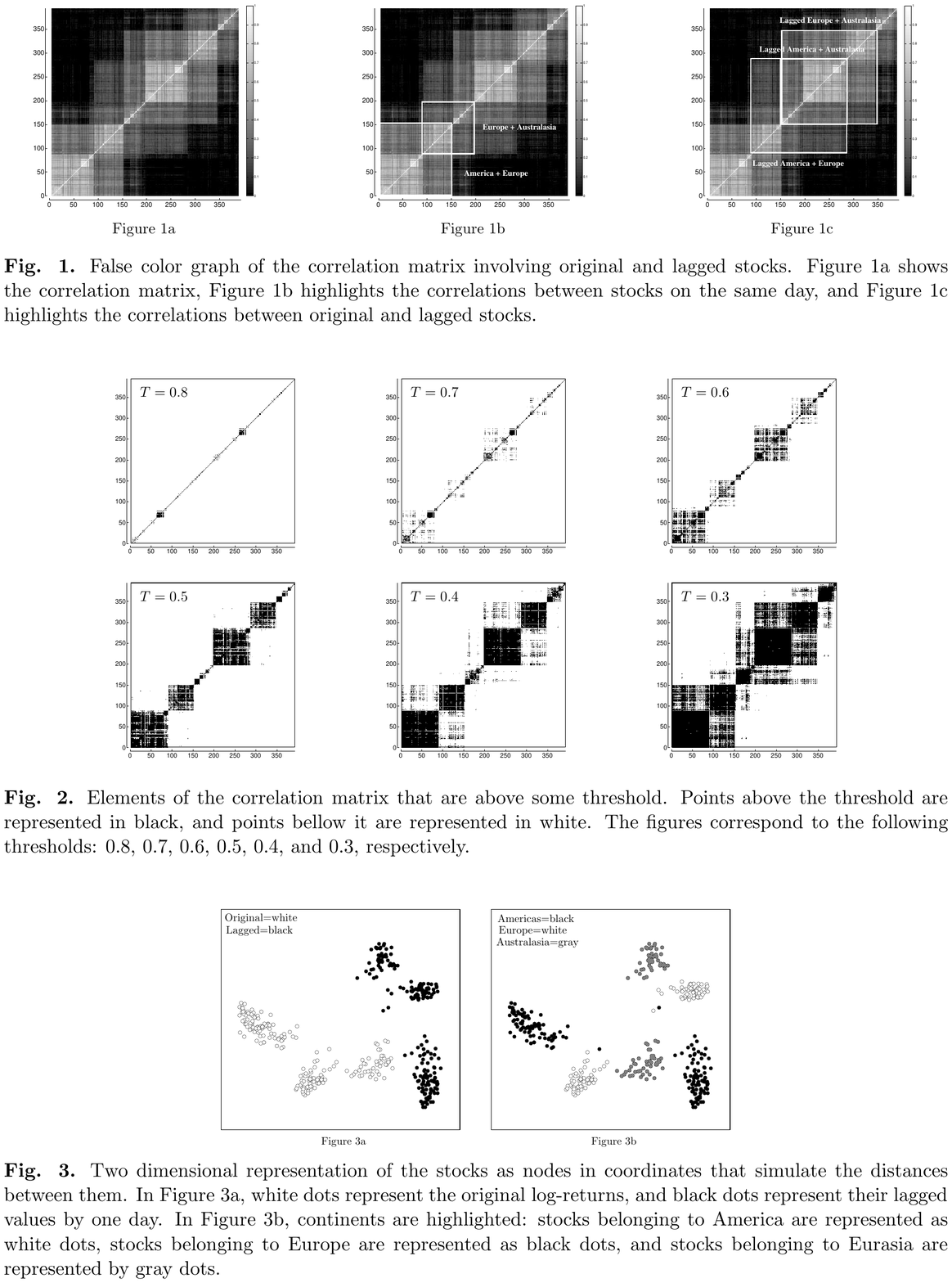}
\end{center}
\end{figure}

\begin{figure}[H]
\begin{center}
\includegraphics[scale=0.85]{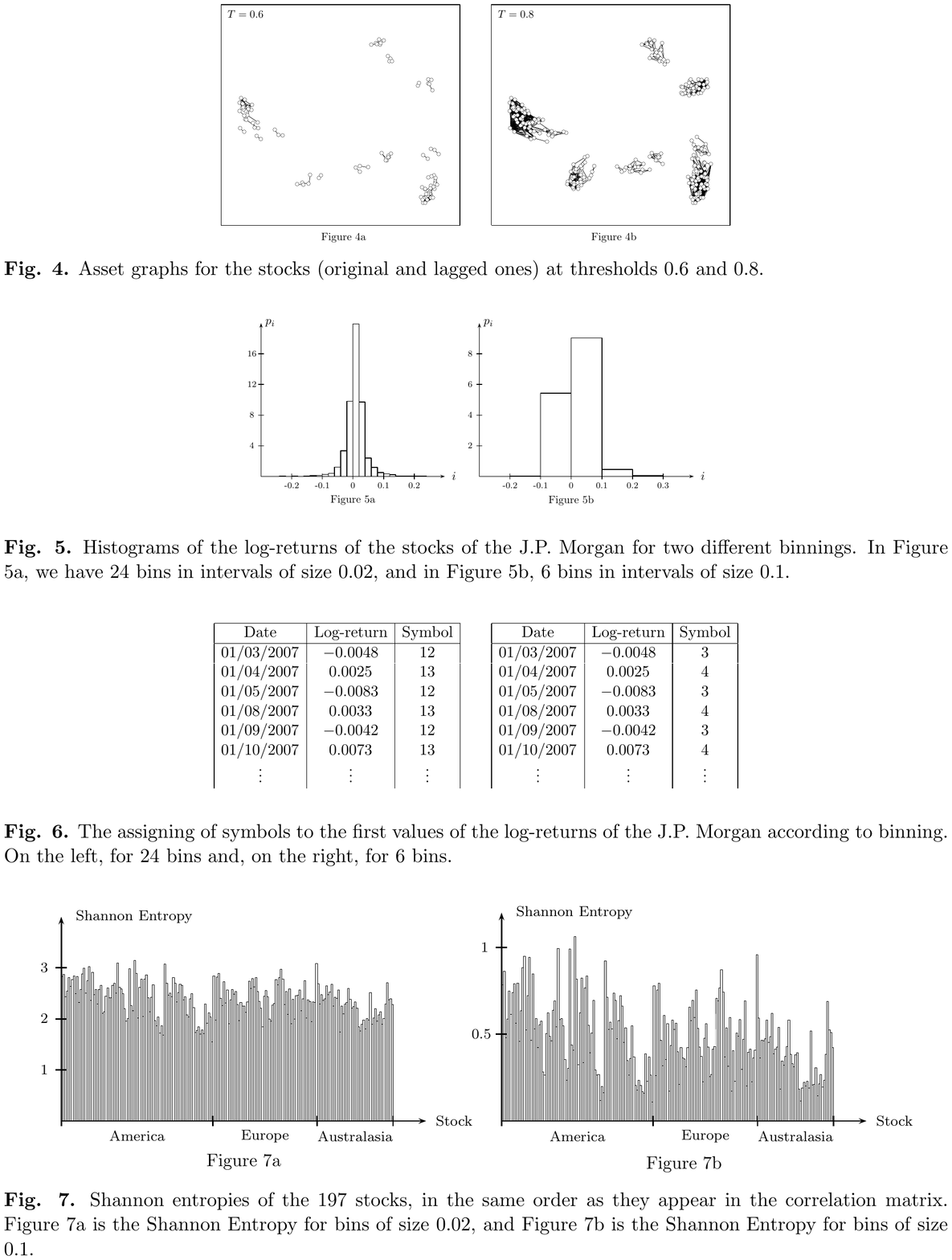}
\end{center}
\end{figure}

\begin{figure}[H]
\begin{center}
\includegraphics[scale=0.85]{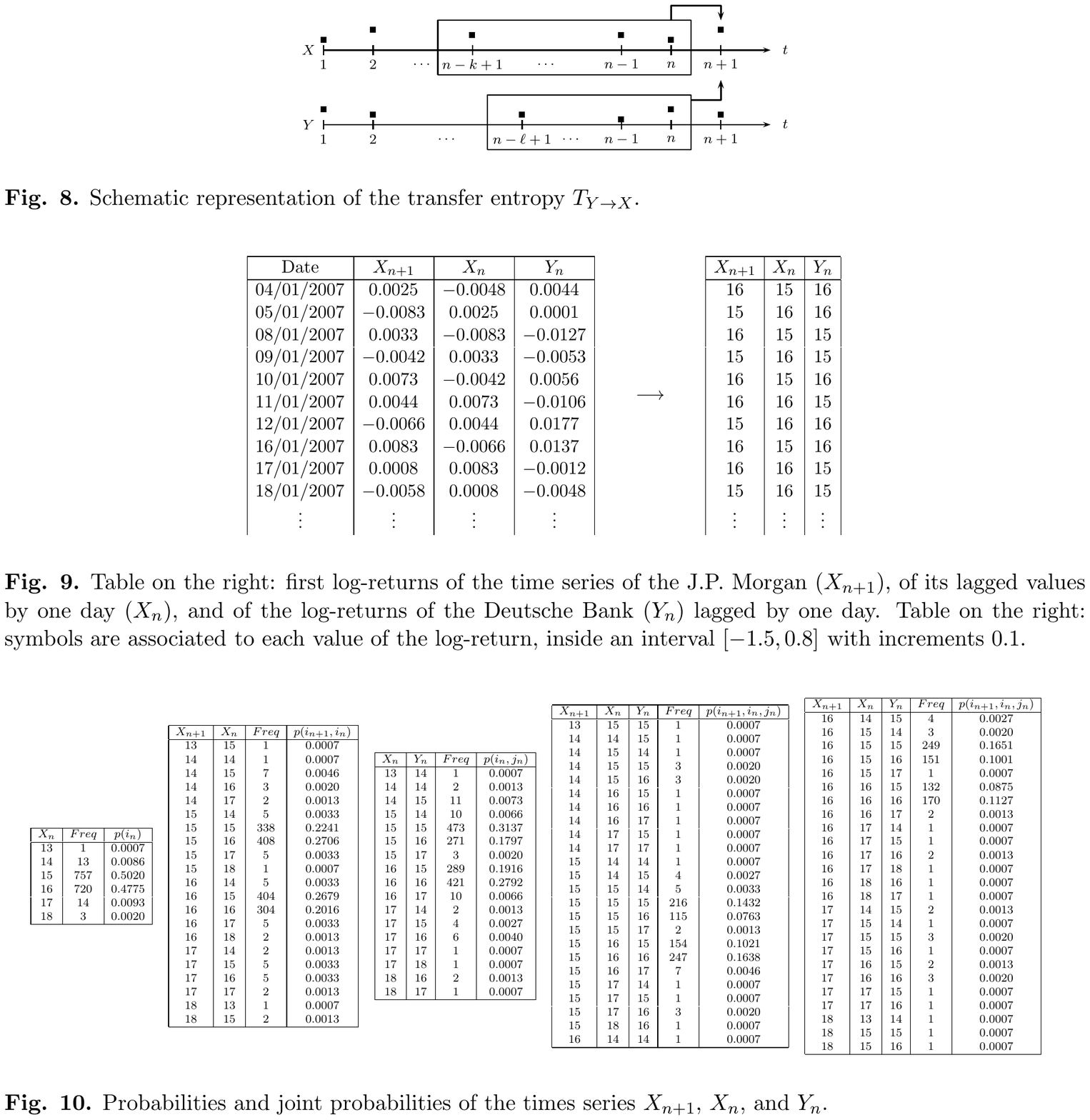}
\end{center}
\end{figure}

\begin{figure}[H]
\begin{center}
\includegraphics[scale=0.85]{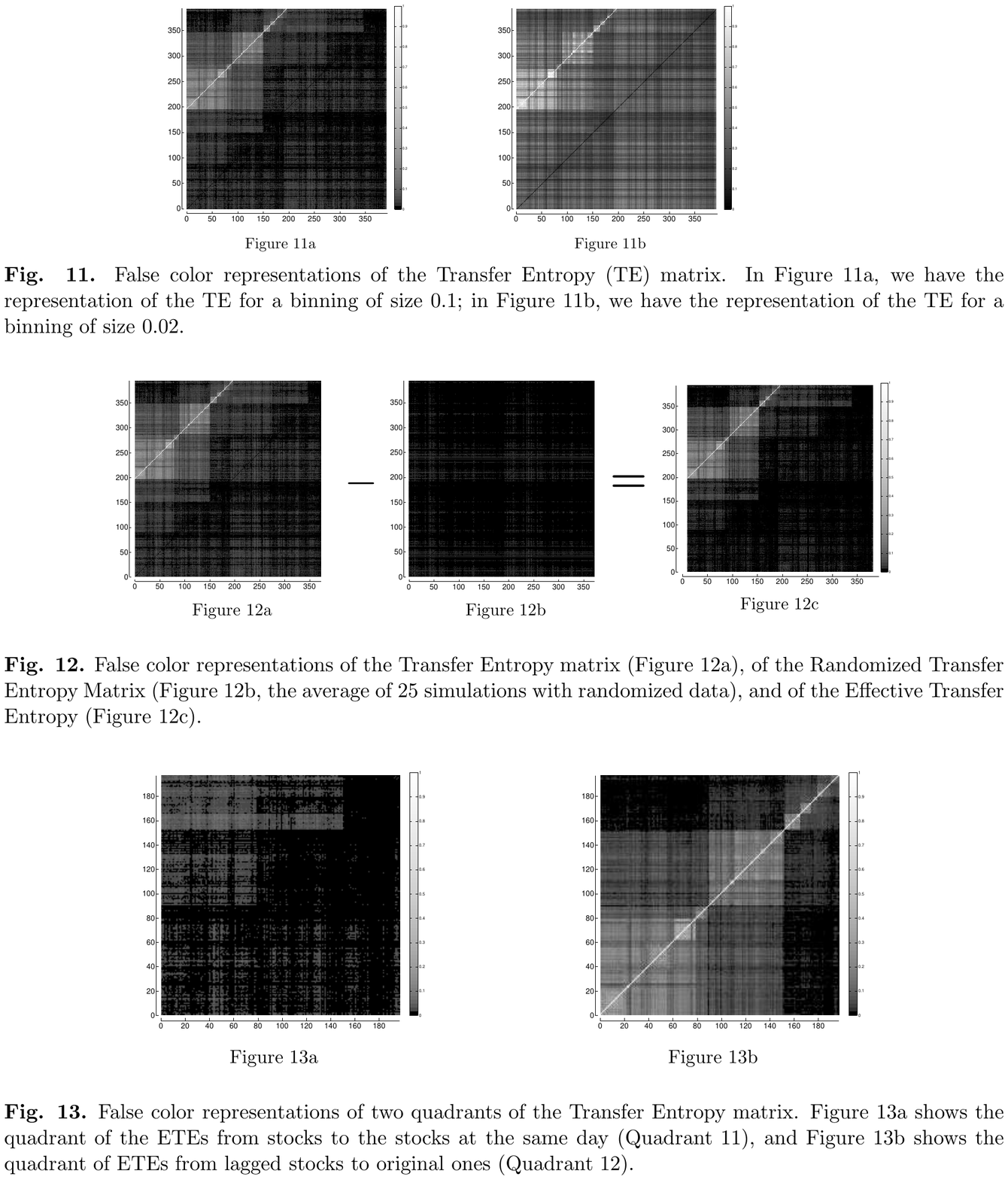}
\end{center}
\end{figure}

\begin{figure}[H]
\begin{center}
\includegraphics[scale=0.85]{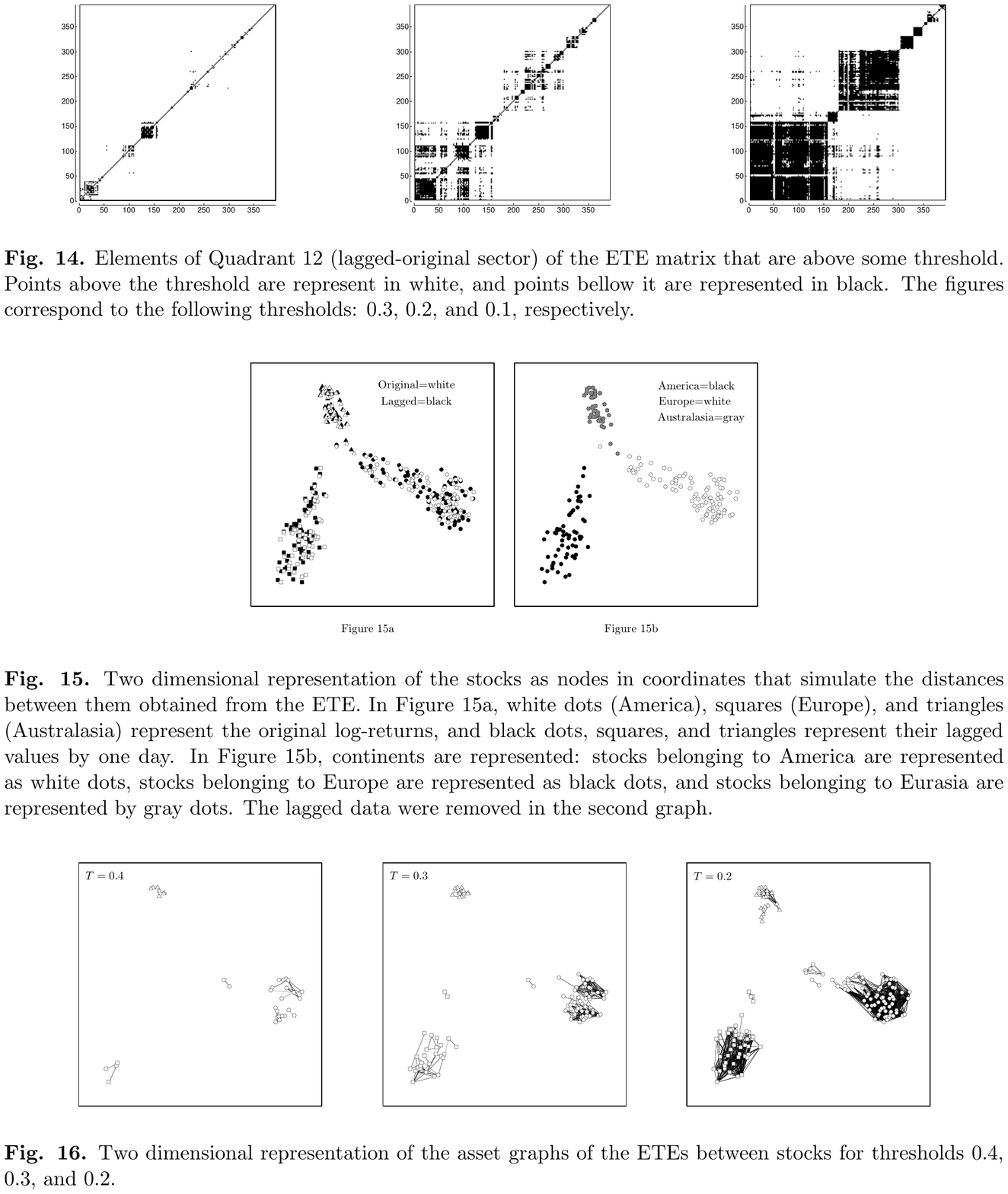}
\end{center}
\end{figure}

\begin{figure}[H]
\begin{center}
\includegraphics[scale=0.85]{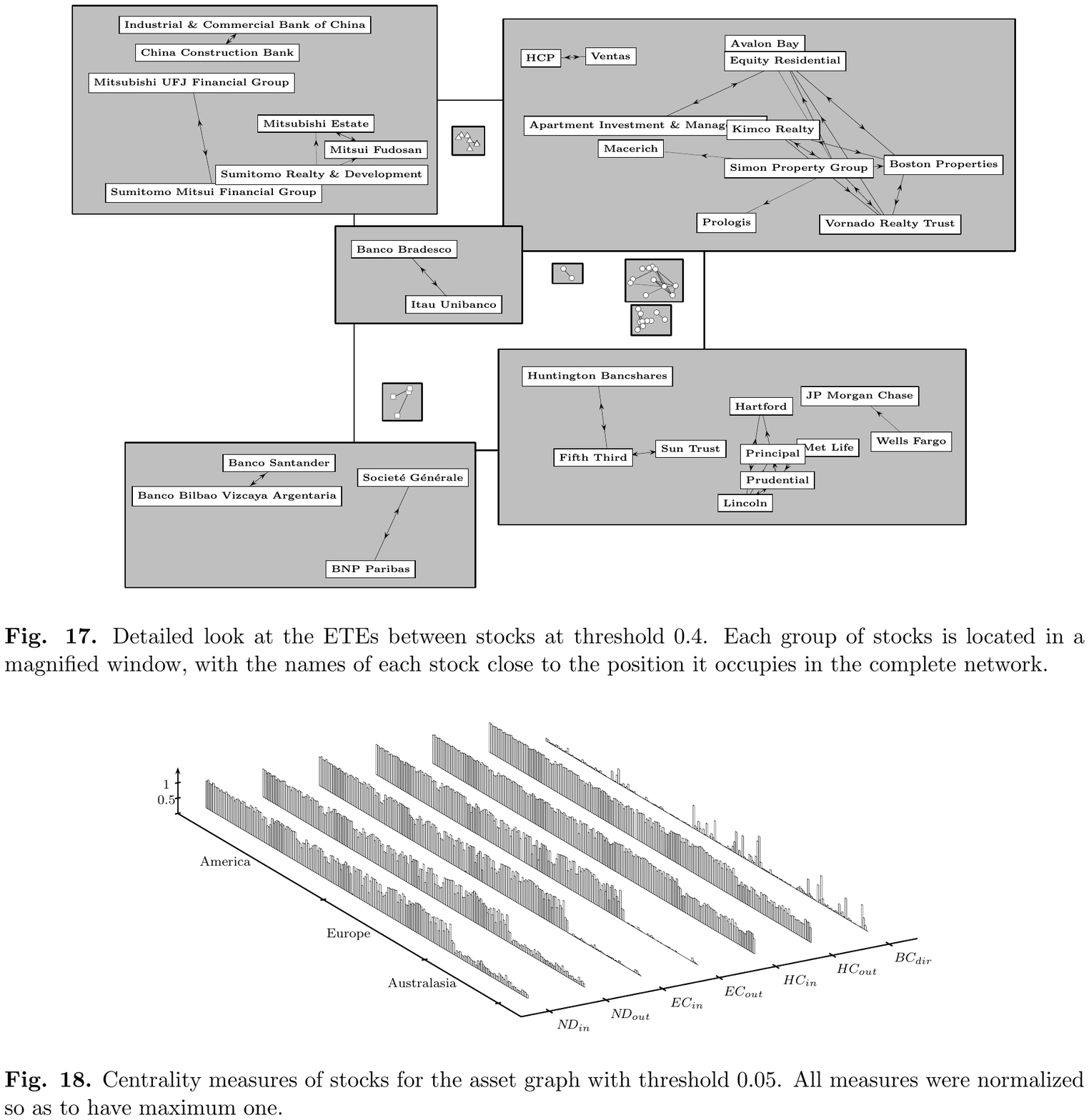}
\end{center}
\end{figure}

\begin{figure}[H]
\begin{center}
\includegraphics[scale=0.85]{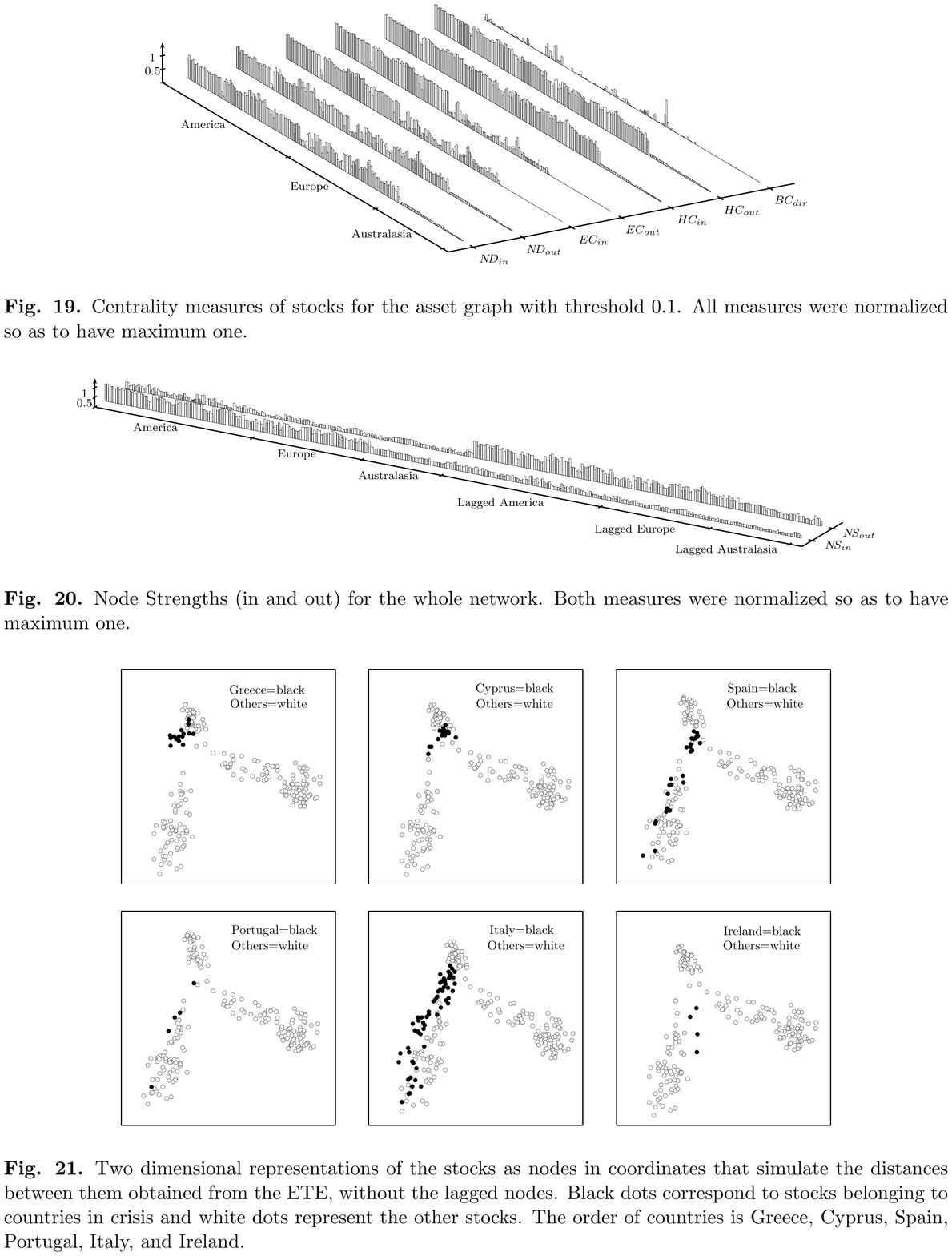}
\end{center}
\end{figure}

\begin{figure}[H]
\begin{center}
\includegraphics[scale=0.85]{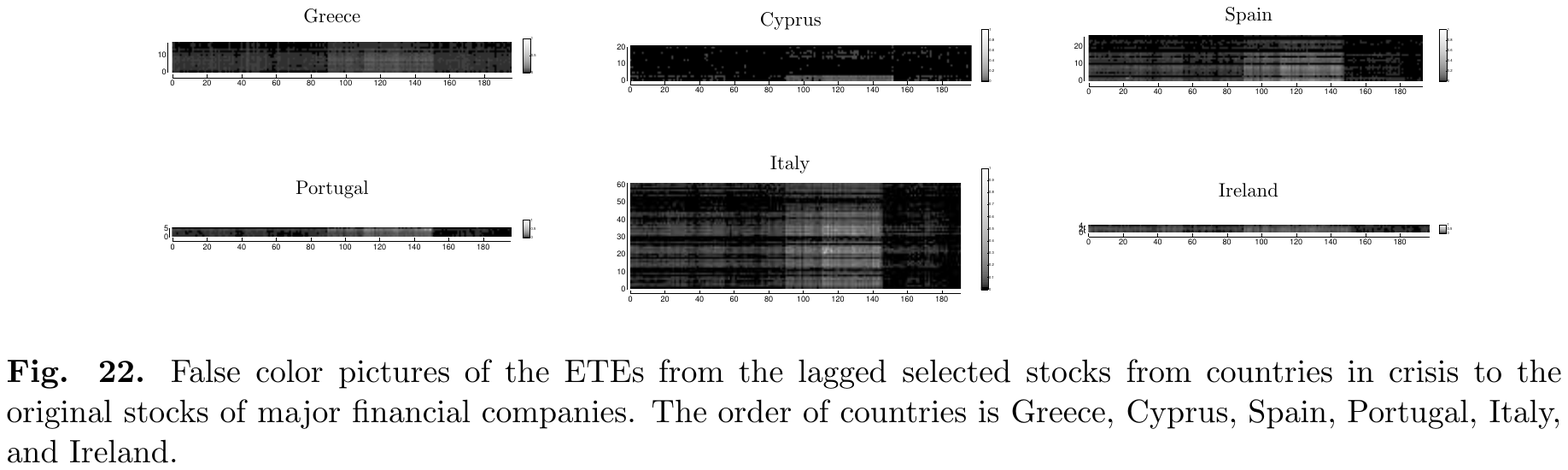}
\end{center}
\end{figure}

\end{document}